%% file: ms.tex
\documentclass[12pt,preprint]{aastex}

\newcommand{\msun}{$M_{\odot}$}

\newcommand{\mjup}{$M_{\rm Jup}$}
\newcommand{\mstar}{$M_{\rm star}$}
\newcommand{\teff}{$T_{\rm eff}$}

\newcommand{\mps}{m s$^{-1}$}
\newcommand{\msini}{$M$ sin $i$}

\newcommand{\rchi}{$\chi_\nu^2$}
\newcommand{\elimit}{$e < 0.8$}

\begin{document}

\title{Fourteen New Companions from the Keck \& Lick Radial Velocity Survey Including Five Brown Dwarf Candidates\altaffilmark{1}}

\author{Shannon G. Patel\altaffilmark{2}, 
Steven S. Vogt\altaffilmark{2},
Geoffrey W. Marcy\altaffilmark{3},
John A. Johnson\altaffilmark{3},
Debra A. Fischer\altaffilmark{4},
Jason T. Wright\altaffilmark{3},
R. Paul Butler\altaffilmark{5}}

\altaffiltext{1}{Based on observations obtained at the W. M. Keck Observatory, which is operated as a scientific partnership among the California Institute of Technology, the University of California and the National Aeronautics and Space Administration. The Observatory was made possible by the generous financial support of the W.M. Keck Foundation.}
\altaffiltext{2}{UCO/Lick Observatory, University of California, Santa Cruz, CA 95064; patel@ucolick.org}
\altaffiltext{3}{Department of Astronomy, University of California, Berkeley, CA 94720-3411}
\altaffiltext{4}{Department of Physics and Astronomy, San Francisco State University, San Francisco, CA 94132}
\altaffiltext{5}{Department of Terrestrial Magnetism, Carnegie Institution of Washington, 5241 Broad Branch Road NW, Washington, DC 20015-1305}

\begin{abstract}
We present radial velocities for 14 stars on the California \& Carnegie Planet Search target list that reveal new companions.  One star, HD 167665, was fit with a definitive Keplerian orbit leading to a minimum mass for the companion of 50.3 \mjup\ at a separation from its host of $\sim$ 5.5 AU.  Incomplete or limited phase coverage for the remaining 13 stars prevents us from assigning to them unique orbital parameters.  Instead, we fit their radial velocities with Keplerian orbits across a grid of fixed values for \msini\ and period, $P$, and use the resulting \rchi\ surface to place constraints on \msini, $P$, and semimajor axis, $a$.  This technique allowed us to restrict \msini\ below the brown dwarf -- stellar mass boundary for an additional 4 companions (HD 150554, HD 8765, HD 72780, HD 74014).  If the combined 5 companions are confirmed as brown dwarfs, these results would comprise the first major catch of such objects from our survey beyond $\sim$ 3 AU.
\end{abstract}


\section{Introduction}

The California \& Carnegie Planet Search (CCPS) has successfully identified over 120 planetary companions to stars in the solar neighborhood with the Doppler technique \citep{butler06}.  While the focus of the survey is the search for planets, the presence of brown dwarfs and low-mass stellar companions in the radial velocity data serves as another avenue for study.  Brown dwarfs in particular have emerged as intriguing research targets as they occupy the region of the mass function between stars and planets with mass boundaries of $\sim$ 80 \mjup\ and $\sim$ 13 \mjup\ respectively.  Unlike stars, which avoid collapse by gas pressure from the high temperatures and densities in their cores that ignite hydrogen fusion, brown dwarfs are supported from collapse by electron degeneracy pressure.  For this reason, the minimum mass necessary to sustain hydrogen burning is taken to be the upper mass limit for brown dwarfs.  The lower mass boundary for brown dwarfs is determined by the deuterium burning limit.  The precise values for these transition masses are not concrete as they depend on parameters such as metallicity.

With their larger imparted reflex velocity on the primary star, brown dwarfs are more amenable to detection than planets.  However, one significant result from the first $\sim 5$ years of our radial velocity survey is the lack of detected companions orbiting within $\sim$ 3 AU of solar-type stars with minimum mass, \msini, in the brown dwarf regime -- a phenomenon seen in earlier radial velocity surveys \citep{campbell88, murdoch93} and commonly referred to as the `brown dwarf desert' \citep{marcy00}.  Planets however, are more routinely found in our survey with their mass function rising towards lower masses \citep{marcy05a}.  Such a dichotomy would seem to indicate distinct formation mechanisms for these two classes of substellar objects.

The lack of brown dwarf orbital companions around Sun-like stars has been studied in detail by several groups.  \citet{grether06} combined data from the major radial velocity surveys for those stars observed over a time baseline of 5 years, resulting in uniform detection of all companions more massive than Jupiter within $\sim$ 3 AU of the host star.  Their results confirm the rising mass function for planetary companions towards lower masses and reveal a rising mass function for stellar companions towards higher mass, effectively leaving a valley around the brown dwarf desert.  

Despite the clear deficit of brown dwarf mass companions in the orbital parameter space currently accessible by the major radial velocity surveys, there are often exceptions.  \citet{nidever02} reported 7 companions with limited radial velocity phase coverage with \msini\ below the hydrogen burning limit.  Since that publication however, subsequent observations leading to more complete phase coverage have pushed at least 1 of those companions into the stellar regime (HD 65430).  \citet{endl04} also report the discovery of a brown dwarf candidate in the desert with minimum mass \msini\ = 26 \mjup, and period $P =$ 798 days.  However, {\em Hipparcos} astrometry measurements do not constrain the companion's mass to be entirely in the brown dwarf regime.

One explanation put forth for the presence of a desert at close separations is the merger of brown dwarfs with their host stars due to orbital migration caused by interactions with an evolving protoplanetary disk \citep{armitage02}.  The theory leaves open the possibility of finding brown dwarfs as companions to solar type stars at larger separations.  While the simulations of \citet{matzner05} also confirm the existence of a desert, they argue that brown dwarfs are not likely to form in disks as suggested by \citet{armitage02}.  The argument removes a wider range of orbital separations available to brown dwarfs and is strengthened by the coronagraphic survey of \citet{mccarthy04} which reveals a desert that extends beyond $\sim$ 1000 AU.

Here we present radial velocities revealing 14 new companions from the CCPS survey.  Some of the inferred companions are examples of potential brown dwarf candidates in the desert.  For one star, HD 167665, we were able to fit a well-determined Keplerian orbit.  The remaining 13 stars lack the phase coverage required to obtain a single set of orbital parameters.  For these targets, we explore \rchi\ for Keplerian fits across a grid of fixed values for \msini\ and $P$, thus providing constraints on these parameters.  This technique is discussed further in \S\ \ref{grid}.  In \S\ \ref{obs} we discuss our observations and technique for radial velocity extraction from the spectra.  Stellar properties and orbital parameters for new companions are presented in \S\ \ref{results}.  We discuss alternative means for constraining these orbits and offer concluding remarks in \S\ \ref{disc}.

\section{Observations} \label{obs}
Three telescopes and two high resolution echelle spectrometers are currently employed by the CCPS in the Northern Hemisphere.  The Keck I 10-m telescope is used in conjunction with HIRES \citep{vogt94}, while either the 3-m Shane or the 0.6-m Coud\'{e} Auxiliary Telescope (CAT) at Lick Observatory feed the Hamilton spectrometer \citep{vogt87}.  The spectrometers provide resolving powers of $R \approx 60,000$ and 55,000 for HIRES and the Hamilton respectively.  The CCPS also uses the AAT and Magellan telescopes for coverage of the Southern Hemisphere.

Our radial velocity extraction technique, discussed at length in \citet{butler96} and summarized below, allows us to obtain high precision relative radial velocities with uncertainties of 2.5--4 \mps\ for the majority of our stars \citep{marcy05b}.  After traversing the optics of the telescope, light from the star passes through a glass cell of iodine gas before entering the slit of the spectrometer.  As a result, thousands of iodine absorption lines are imprinted on the observed stellar spectrum thus serving as a wavelength fiducial.  A model stellar spectrum consisting of a high signal-to-noise template stellar spectrum multiplied by the predetermined transmission function of the iodine cell is then convolved with the PSF of the spectrometer and fit to the observed stellar spectrum.  Among the many parameters to the fit is the Doppler shift of the template stellar spectrum.

The target list of stars for our survey is drawn from the \textit{Hipparcos} catalogue \citep{esa97} with the criteria that they have $B - V > 0.55$.  For a detailed discussion on our selection criteria see \citet{marcy05b}.  Select stellar properties for the majority of our program stars can be found in \citet{wright04}

Observations for the 14 stars with companions presented here were obtained as far back as 1996.  Because these companions are more massive than the planets found in our survey, they impart a larger radial velocity amplitude on the primary star.  As a consequence, we typically obtain spectra for the host star with shorter integration times resulting in larger photon-limited velocity precision ($\sim$ 1.5--15 \mps).

\section{Results} \label{results}
\subsection{Stellar Properties}

Properties for stars with companions presented in this paper are given in Table \ref{stars}.  Columns (1) and (2) indicate the HD and \textit{Hipparcos} catalogue identification numbers respectively.  The spectral type in column (3) is that given by SIMBAD.  When available, the luminosity class was also taken from SIMBAD, otherwise the derived stellar parameters given in the table were used with \citet{cox00} to provide a best guess.  All of the host stars are dwarfs with the exception of four subgiants (HD 211681, HD 215578, HD 29461, HD 8765).  The distance ($d$) and absolute visual magnitude ($M_V$), given in columns (4) and (5) respectively, are taken from \textit{Hipparcos}.  Masses for the host stars in column (6) were obtained from \citet{takeda07} with the exception of HD 215578, whose mass was obtained from \citet{girardi02}.  The next four columns indicate surface temperature (7), surface gravity (8), metallicity (9), and rotation velocity (10), which were obtained from \citet{valenti05}.  The primary observing facility is given in column (11).

\subsection{HD 167665} \label{hd167665}

\begin{figure}
\plotone{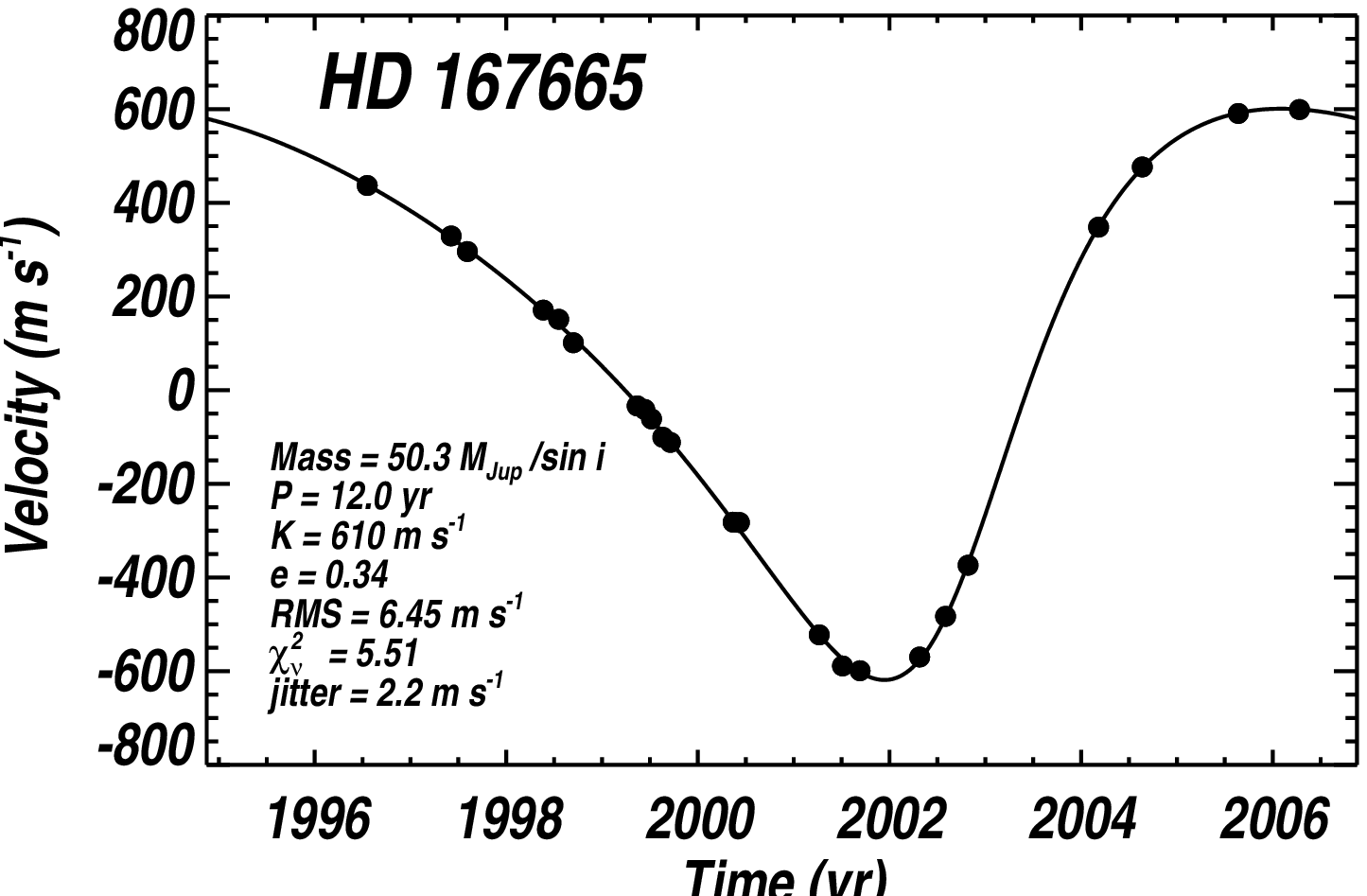}
\caption{Velocity vs. time for HD 167665 (dots).  The solid line is the best-fit Keplerian orbit with $P$ = 12.0 yr, $K$ = 610 \mps, $e$ = 0.34, and \msini\ = 50.3 \mjup. } \label{rv167665}
\end{figure}

Orbital parameters for HD 167665 are given in Table \ref{orbital}.  These were derived from times of observations, radial velocities and instrumental uncertainties that are given in Table \ref{trverr167665}.  Radial velocities with best-fit Keplerian orbit are shown in Figure \ref{rv167665}.  Instrumental uncertainties on the individual radial velocity measurements are also plotted in this figure but are in most cases too small to see at the given scale.  A Levenberg-Marquardt algorithm was implemented to determine the best Keplerian fit to the observed data.  Uncertainties for the orbital parameters were determined as in \citet{marcy05b}.  The residuals from the best-fit Keplerian orbit were treated as samples of the uncertainty due to the combined Doppler error and photospheric jitter.  One hundred mock velocity sets were then created where a new mock velocity was constructed by adding the original observed velocity to a randomly-picked residual.  The final uncertainties for the orbital parameters as reported in Table \ref{orbital} are given by the standard deviation of the best-fit parameters to the 100 mock velocity sets.  The uncertainty for \msini\ was calculated by propagating the errors for the relevant orbital parameters through the mass function equation.  Note that the quoted uncertainty for \msini\ does not reflect the uncertainty associated with the host mass, which is typically $\sim$ 5--10\%.  The error for the minimum semimajor axis, $a_{\rm min}$, was computed by propagating the relevant errors through Kepler's 3rd Law.

The companion to HD 167665 orbits with a period of $\sim$ 12 yr at a separation from its host of $\sim$ 5.5 AU.  It has a minimum mass firmly established in the brown dwarf regime, with \msini\ = 50.3 \mjup.  The true mass depends on the inclination of the orbit with higher inclinations leading to larger true masses.  If we assume the orbit is randomly oriented on the unit sphere, there is a 79\% chance that the inclination favors an object in the brown dwarf mass range.  We take 82 \mjup\ as the mass boundary between stars and brown dwarfs by interpolating between tracks in \citet{burrows97}.

\subsection{Stars with Limited Phase Coverage} \label{grid}

Thirteen of the 14 stars presented in this paper were not observed long enough to realize a full orbit.  Any attempt to present a single set of orbital parameters for them would be misleading as any number of a large family of solutions could result in an acceptable fit.  However, if the radial velocity data shows enough curvature it may be possible to put limits on the period ($P$) and minimum mass (\msini).  Following the analysis in \citet{wright07}, we examine the reduced chi-square statistic, \rchi, resulting from Keplerian fits with fixed values of \msini\ and $P$.  Typically the semiamplitude, $K$, is the parameter of interest when detecting spectroscopic companions.  In order to place constraints on the minimum mass   we transform from ($P$, $e$, $K$) to ($P$, $e$, \msini) coordinates through the mass function equation, $f(M)$:
\begin{equation}
\label{massfn}
  f(M) = \frac{M^3 \sin^3{i}}{(M+M_*)^2} =
  \frac{P K^3(1-e^2)^{\frac{3}{2}}}{2 \pi G}
\end{equation}

For each of our limited phase coverage targets, we create a grid of values for \msini\ and $P$, both in log-step.  These are the fixed values with which the Keplerian fitting routine is executed while allowing $e$, $\omega$, $T_{\rm P}$ and $\gamma$ to vary.  Again, we use a Levenberg-Marquardt algorithm to arrive at a solution, which is in almost all cases the global minimum given the fewer free parameters to the fit and because of the velocity precision at these large amplitudes.  A smoothly varying \rchi\ surface on the \msini\ -- $P$ grid provides further evidence that we are tracing the global minimum.

\begin{figure}
\epsscale{1}
\plottwo{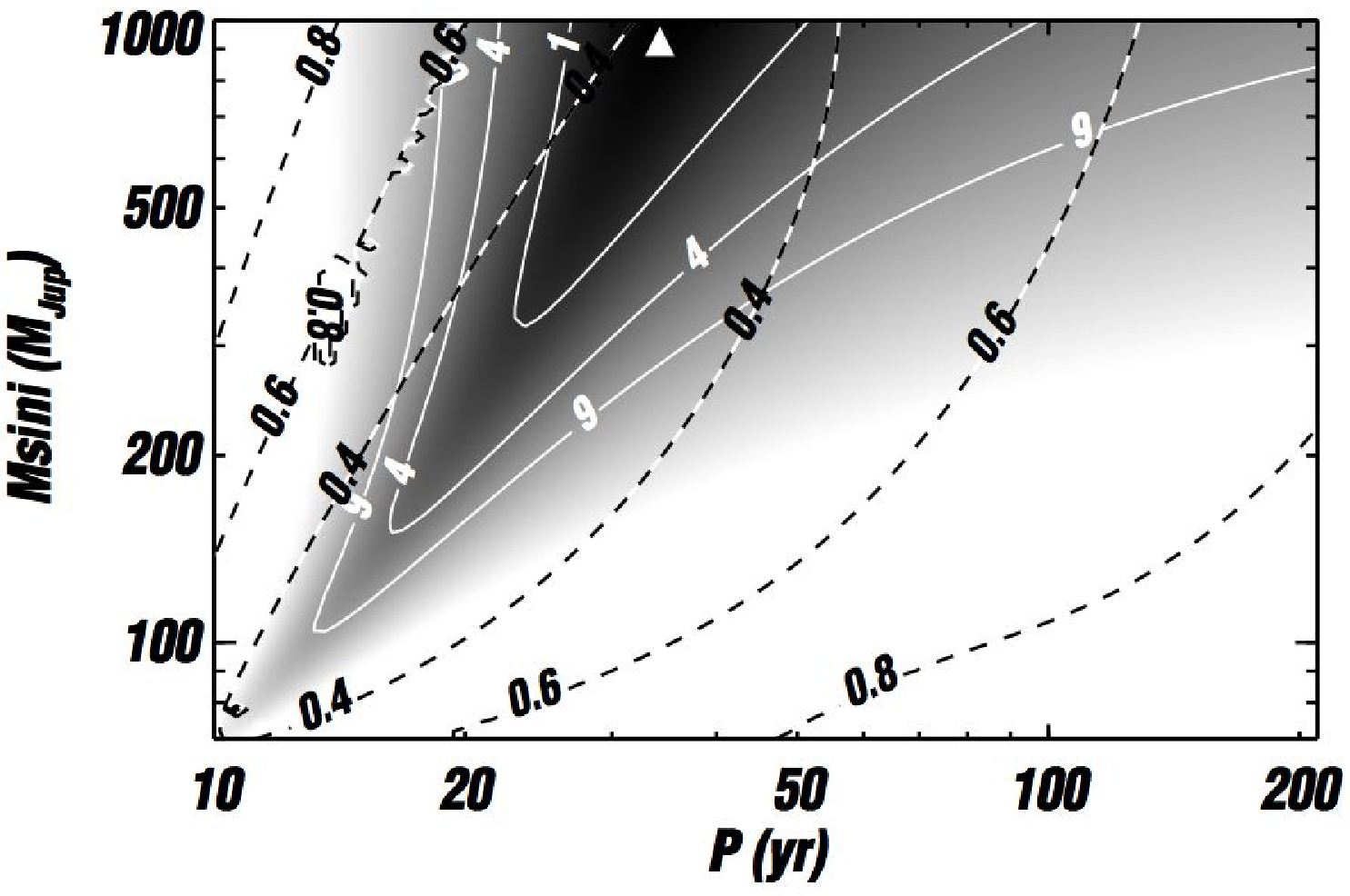}{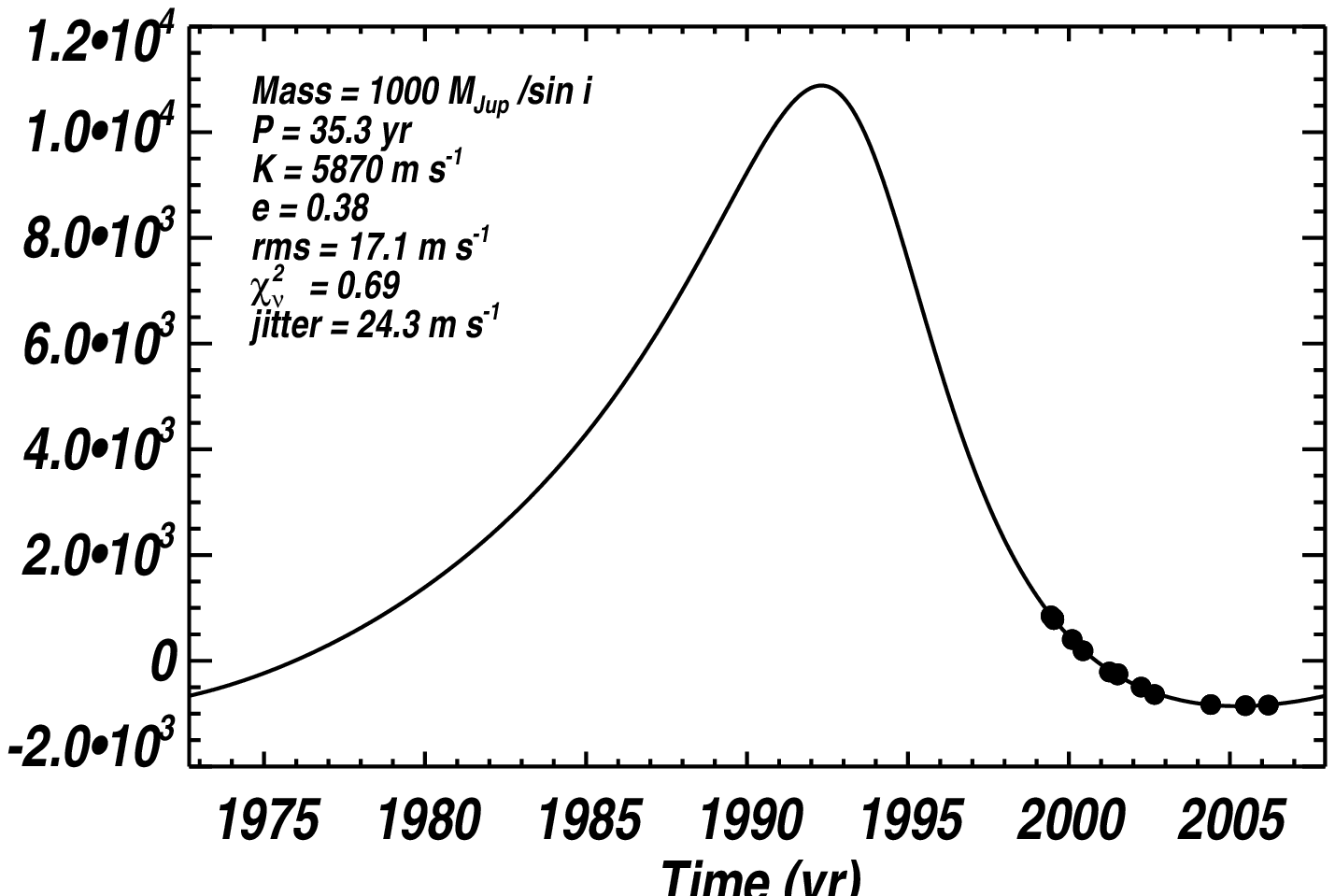}
\caption{ {\em Left}: Contours of $\chi_\nu^2$ in $M$ sin $i$ -- $P$ space for HD 142229.  The solid contour lines represent increases from the best-fit ($\chi_{\nu, {\rm min}}^2$) of 1, 4, and 9.  The location of the best-fit on the $M$ sin $i$ -- $P$ grid is indicated by the arrow on the upper edge at 1000 $M_{\rm Jup}$ and 35.3 yr, and its corresponding Keplerian orbit is plotted on the right.  Contours of eccentricity are shown as dashed lines.  The $\chi_\nu^2$ = $\chi_{\nu, {\rm min}}^2 + 4 $ contour limits the minimum mass and period to: $M$ sin $i$ $(M_{\rm Jup}) >153$, $P$ (yr) $>16.4$.  {\em Right}: Velocity vs. time for HD 142229 (dots).  The solid line is the Keplerian orbit for the best-fit ($\chi_{\nu, {\rm min}}^2$) from the $M$ sin $i$ -- $P$ grid.} \label{plot142229}
\end{figure}

\begin{figure}
\epsscale{1}
\plottwo{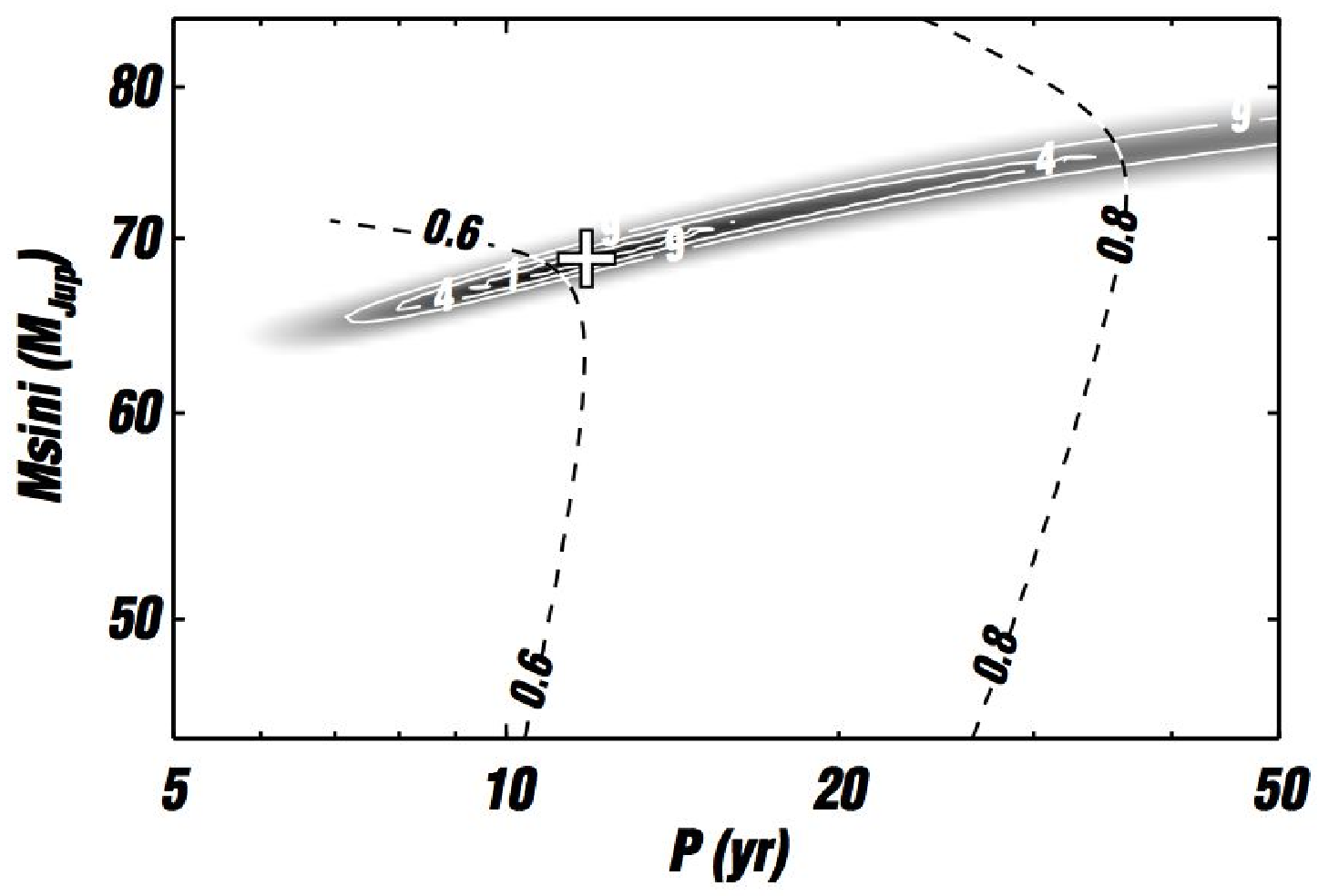}{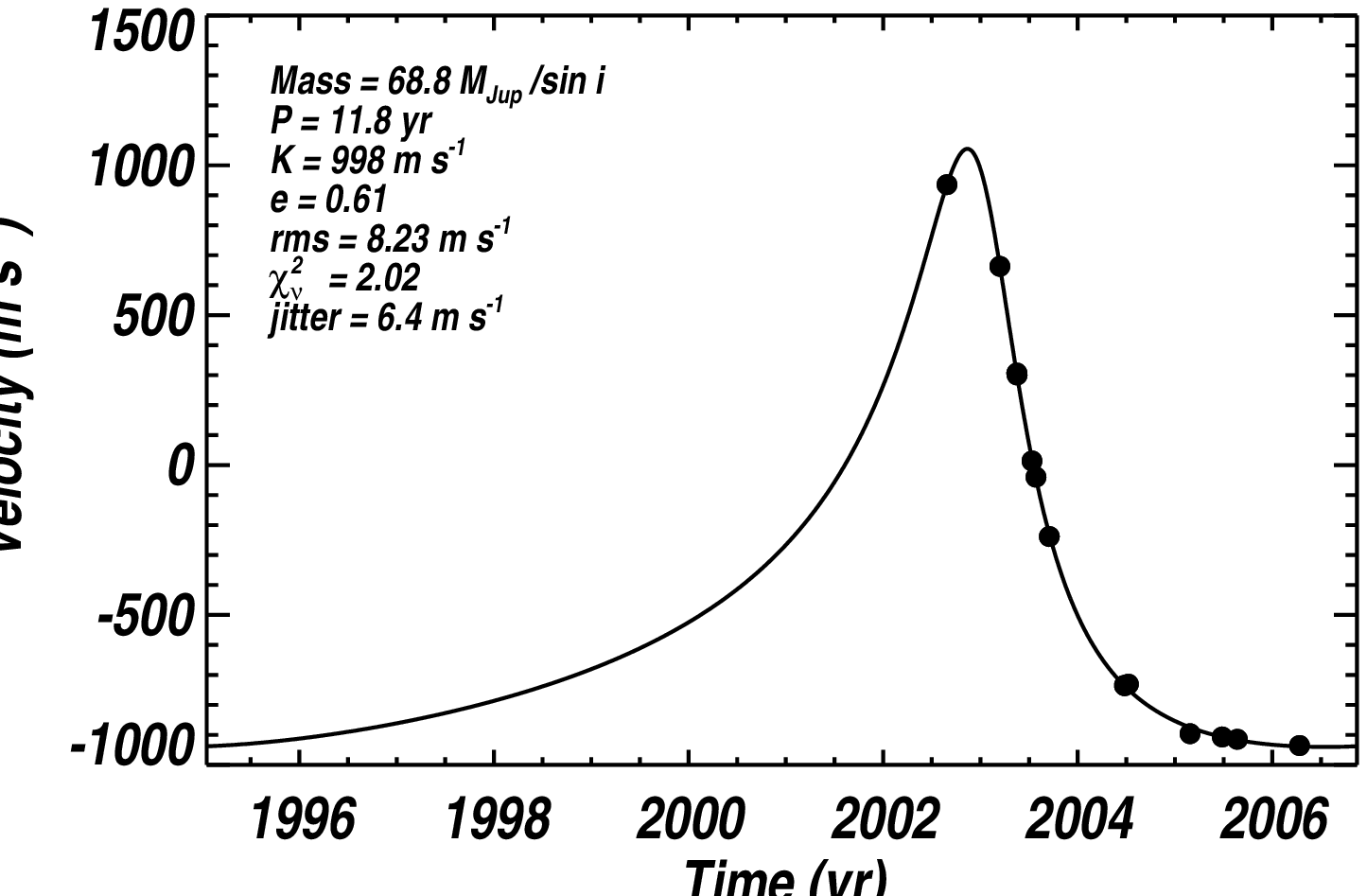}
\caption{ {\em Left}: Contours of $\chi_\nu^2$ in $M$ sin $i$ -- $P$ space for HD 150554.  The solid contour lines represent increases from the best-fit ($\chi_{\nu, {\rm min}}^2$) of 1, 4, and 9.  The location of the best-fit on the $M$ sin $i$ -- $P$ grid is indicated with a cross at 68.8 $M_{\rm Jup}$ and 11.8 yr, and its corresponding Keplerian orbit is plotted on the right.  Contours of eccentricity are shown as dashed lines.  The $\chi_\nu^2$ = $\chi_{\nu, {\rm min}}^2 + 4 $ contour and $e < 0.8$ constraint limits the minimum mass and period to: $65.7 < M$ sin $i$ $(M_{\rm Jup}) < 75.2$, $8.0 < P$ (yr) $< 33.7$.  {\em Right}: Velocity vs. time for HD 150554 (dots).  The solid line is the Keplerian orbit for the best-fit ($\chi_{\nu, {\rm min}}^2$) from the $M$ sin $i$ -- $P$ grid.} \label{plot150554}
\end{figure}

\begin{figure}
\epsscale{1}
\plottwo{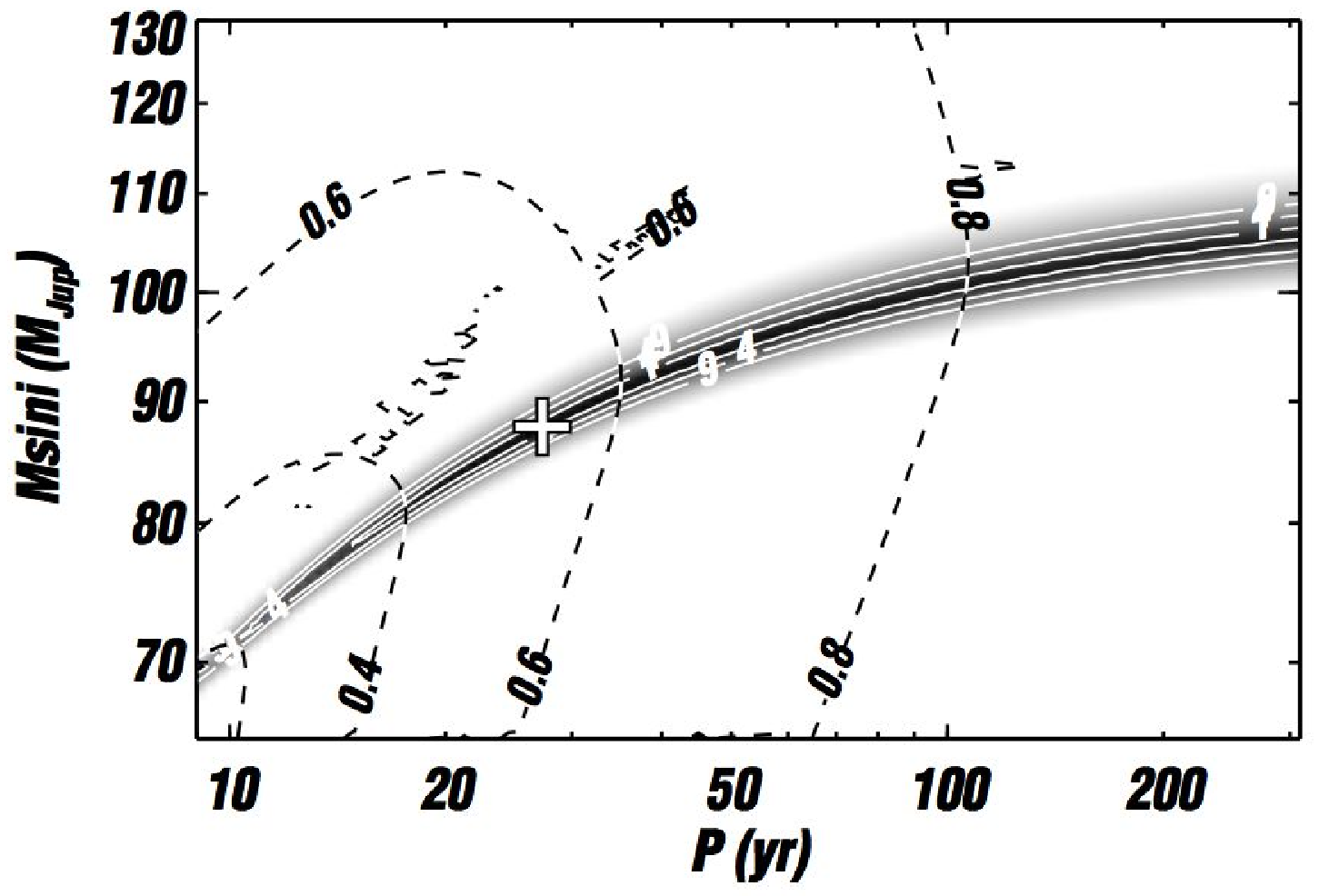}{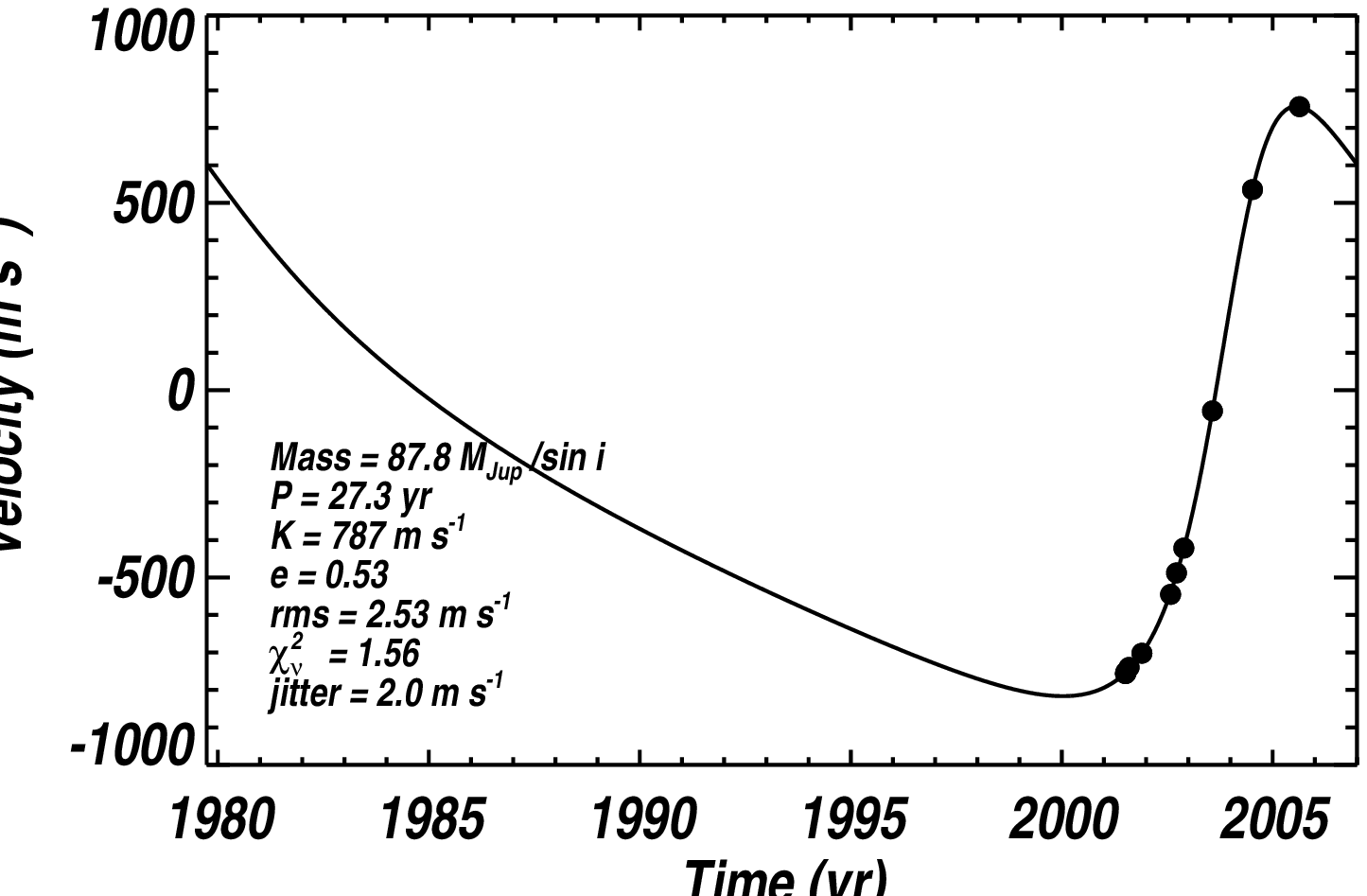}
\caption{ {\em Left}: Contours of $\chi_\nu^2$ in $M$ sin $i$ -- $P$ space for HD 211681.  The solid contour lines represent increases from the best-fit ($\chi_{\nu, {\rm min}}^2$) of 1, 4, and 9.  The location of the best-fit on the $M$ sin $i$ -- $P$ grid is indicated with a cross at 87.8 $M_{\rm Jup}$ and 27.3 yr, and its corresponding Keplerian orbit is plotted on the right.  Contours of eccentricity are shown as dashed lines.  The $\chi_\nu^2$ = $\chi_{\nu, {\rm min}}^2 + 4 $ contour and $e < 0.8$ constraint limits the minimum mass and period to: $71.7 < M$ sin $i$ $(M_{\rm Jup}) < 102$, $10.4 < P$ (yr) $< 106$.  {\em Right}: Velocity vs. time for HD 211681 (dots).  The solid line is the Keplerian orbit for the best-fit ($\chi_{\nu, {\rm min}}^2$) from the $M$ sin $i$ -- $P$ grid.} \label{plot211681}
\end{figure}

\begin{figure}
\epsscale{1}
\plottwo{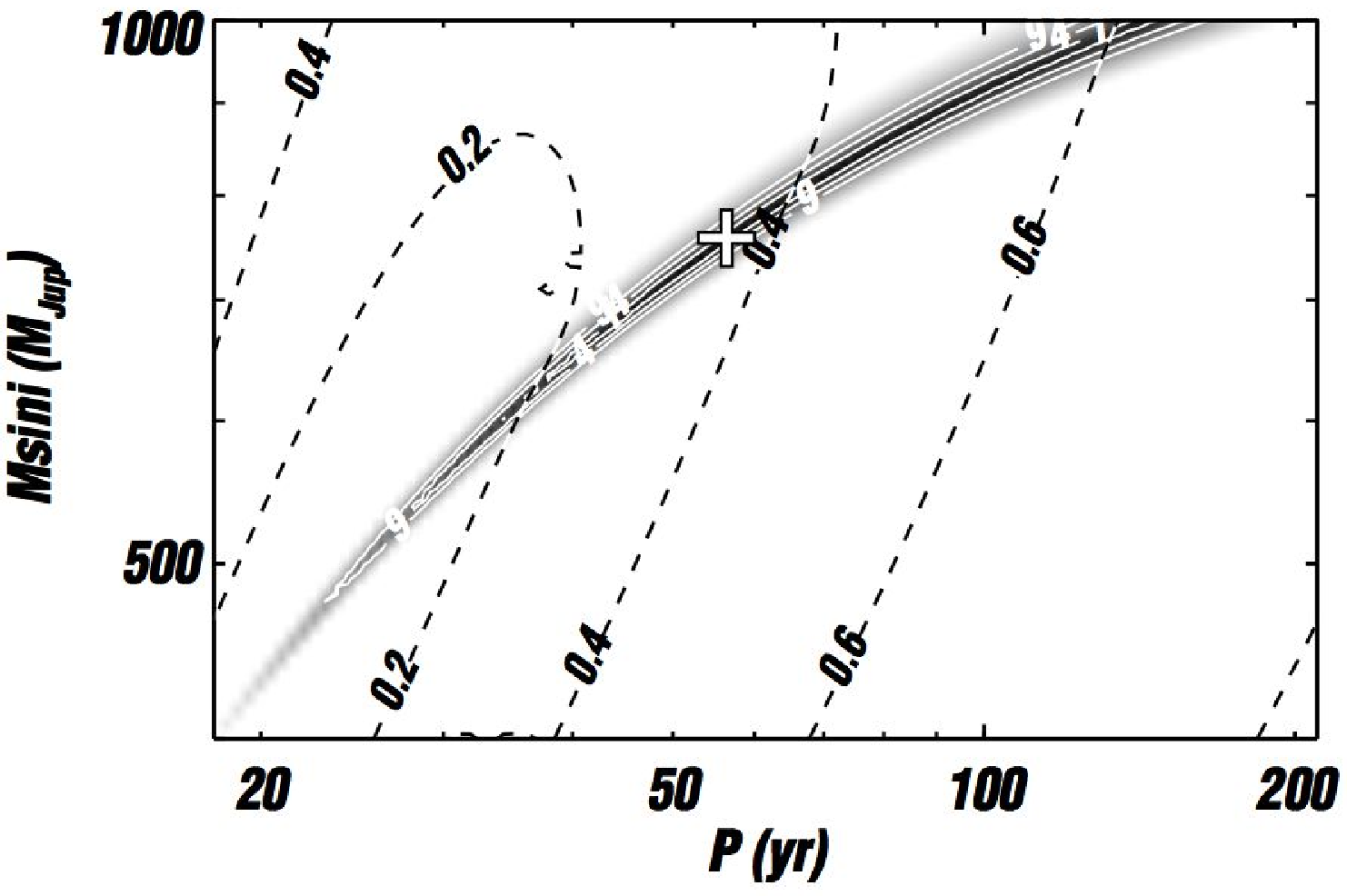}{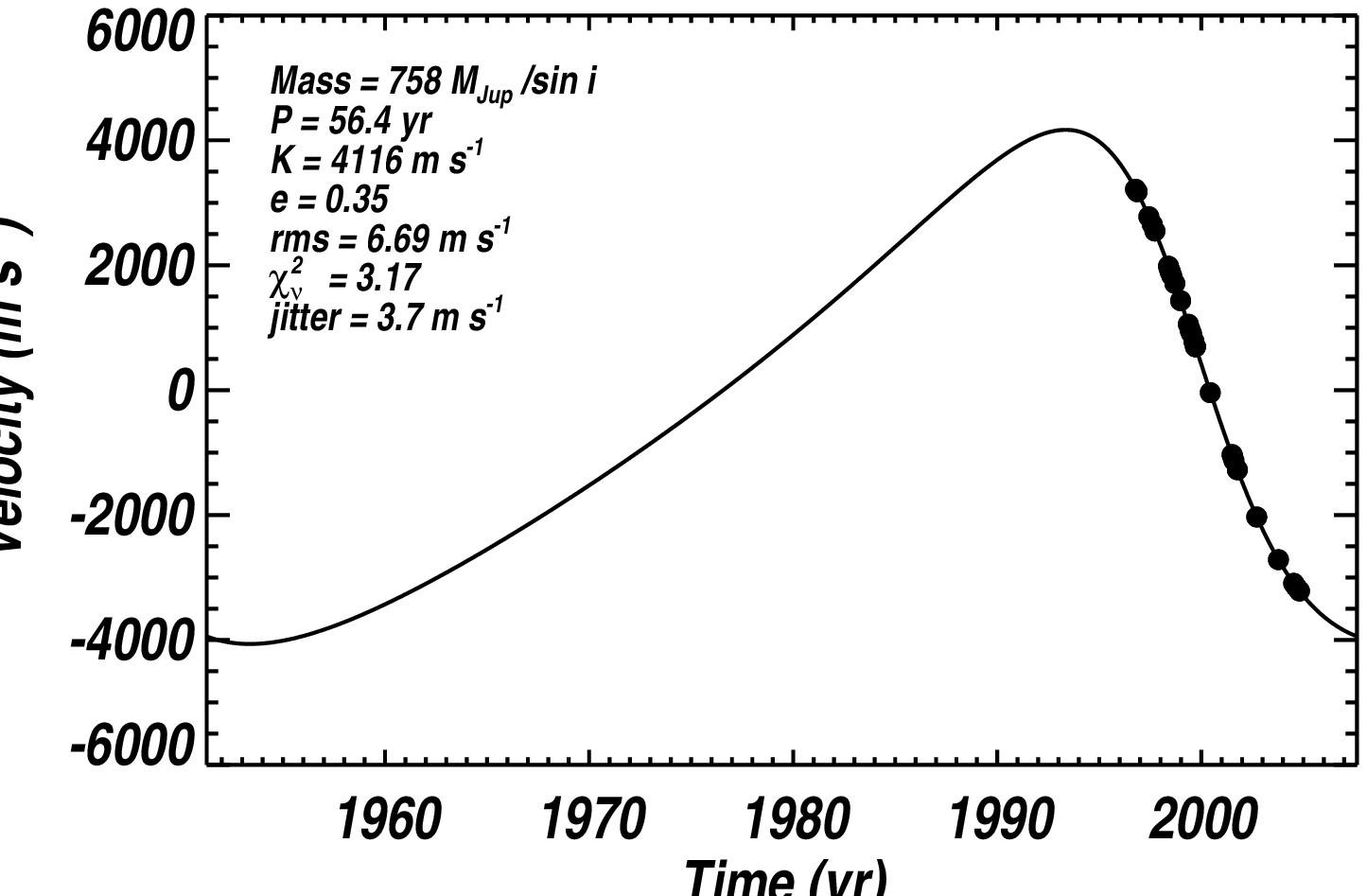}
\caption{ {\em Left}: Contours of $\chi_\nu^2$ in $M$ sin $i$ -- $P$ space for HD 215578.  The solid contour lines represent increases from the best-fit ($\chi_{\nu, {\rm min}}^2$) of 1, 4, and 9.  The location of the best-fit on the $M$ sin $i$ -- $P$ grid is indicated with a cross at 758 $M_{\rm Jup}$ and 56.4 yr, and its corresponding Keplerian orbit is plotted on the right.  Contours of eccentricity are shown as dashed lines.  The $\chi_\nu^2$ = $\chi_{\nu, {\rm min}}^2 + 4 $ contour limits the minimum mass and period to: $M$ sin $i$ $(M_{\rm Jup}) >523$, $P$ (yr) $>26.8$.  {\em Right}: Velocity vs. time for HD 215578 (dots).  The solid line is the Keplerian orbit for the best-fit ($\chi_{\nu, {\rm min}}^2$) from the $M$ sin $i$ -- $P$ grid.} \label{plot215578}
\end{figure}

\begin{figure}
\epsscale{1}
\plottwo{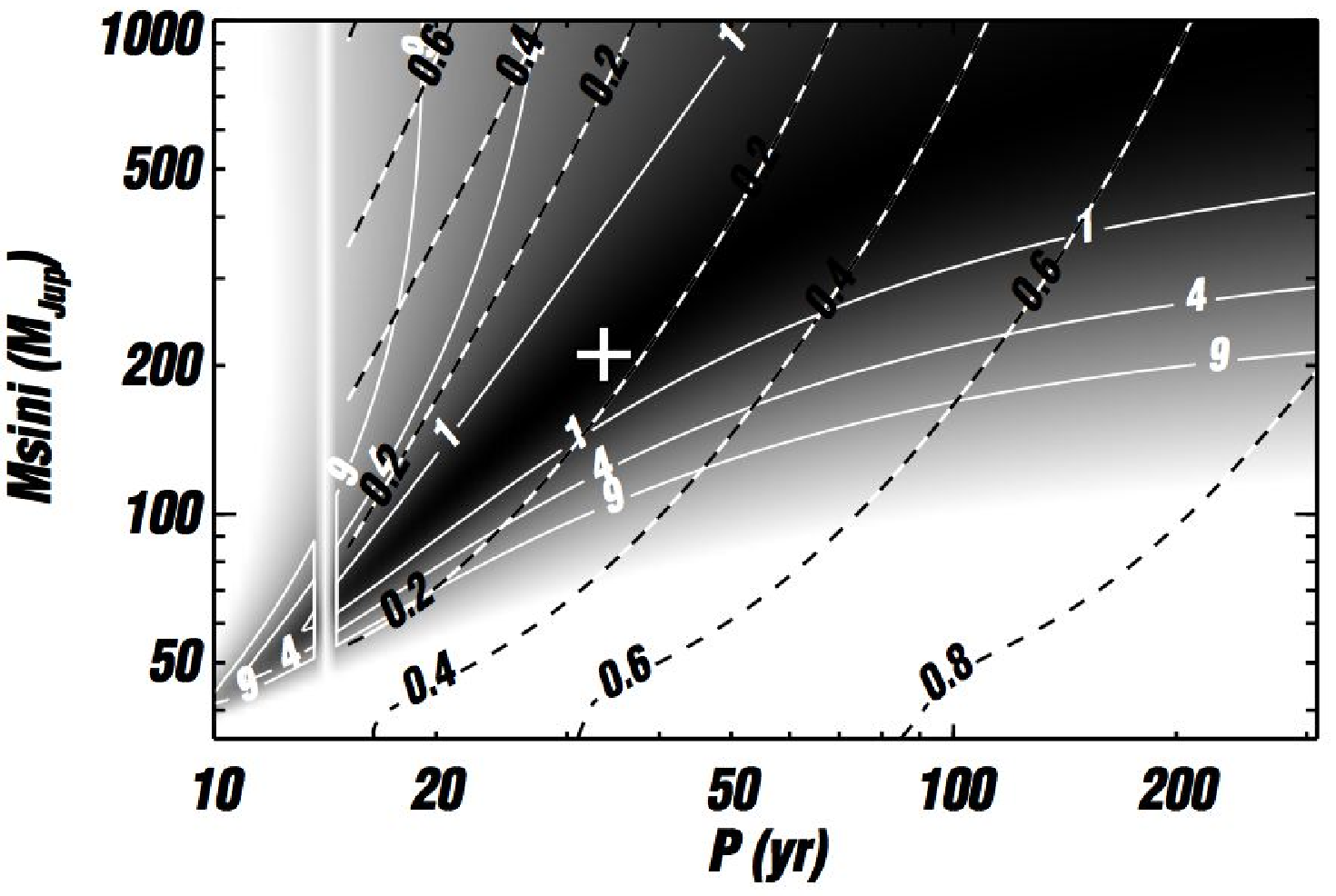}{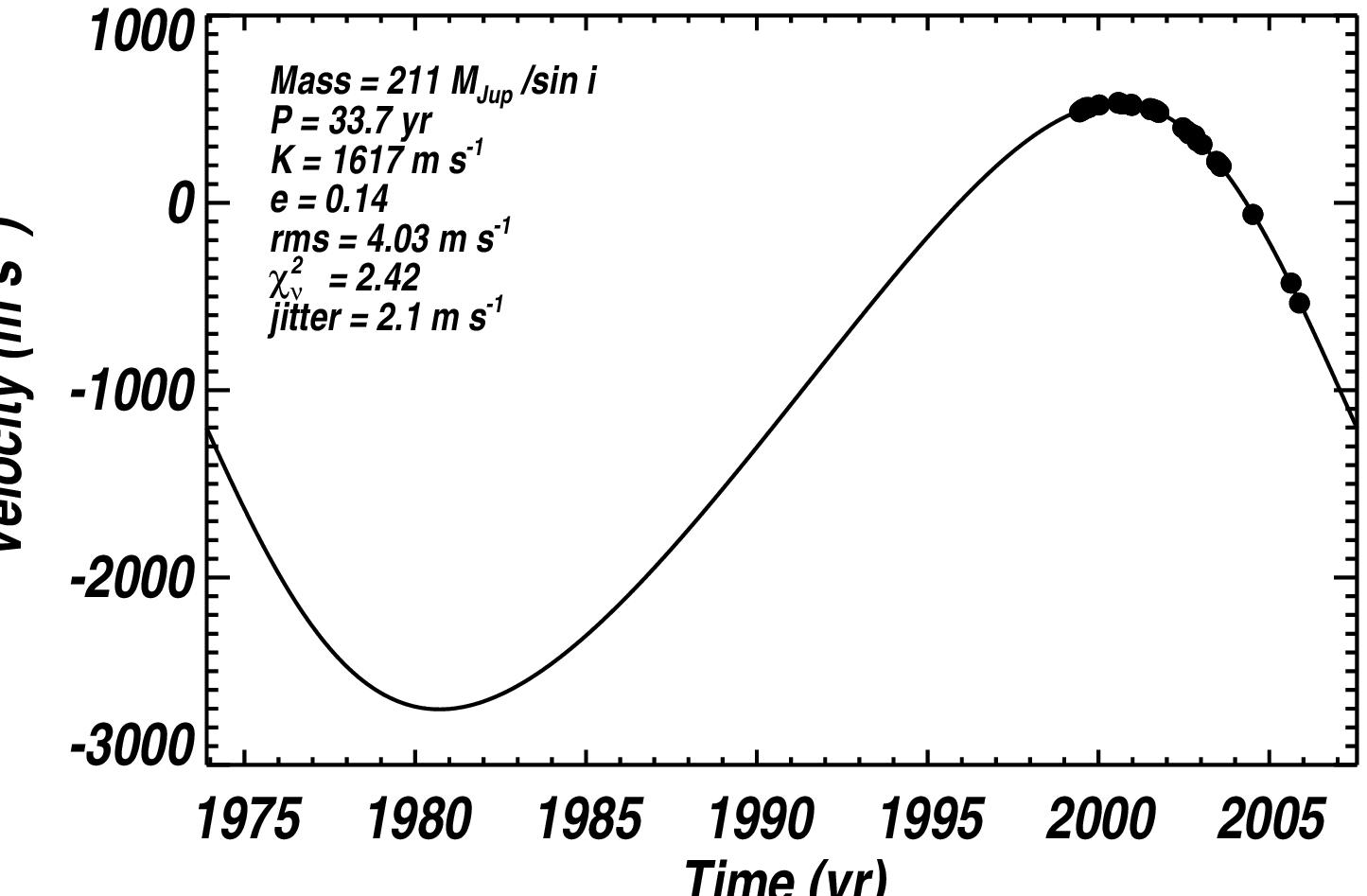}
\caption{ {\em Left}: Contours of $\chi_\nu^2$ in $M$ sin $i$ -- $P$ space for HD 217165.  The solid contour lines represent increases from the best-fit ($\chi_{\nu, {\rm min}}^2$) of 1, 4, and 9.  The location of the best-fit on the $M$ sin $i$ -- $P$ grid is indicated with a cross at 211 $M_{\rm Jup}$ and 33.7 yr, and its corresponding Keplerian orbit is plotted on the right.  Contours of eccentricity are shown as dashed lines.  The $\chi_\nu^2$ = $\chi_{\nu, {\rm min}}^2 + 4 $ contour limits the minimum mass and period to: $M$ sin $i$ $(M_{\rm Jup}) >45.9$, $P$ (yr) $>11.1$.  {\em Right}: Velocity vs. time for HD 217165 (dots).  The solid line is the Keplerian orbit for the best-fit ($\chi_{\nu, {\rm min}}^2$) from the $M$ sin $i$ -- $P$ grid.} \label{plot217165}
\end{figure}

\begin{figure}
\epsscale{1}
\plottwo{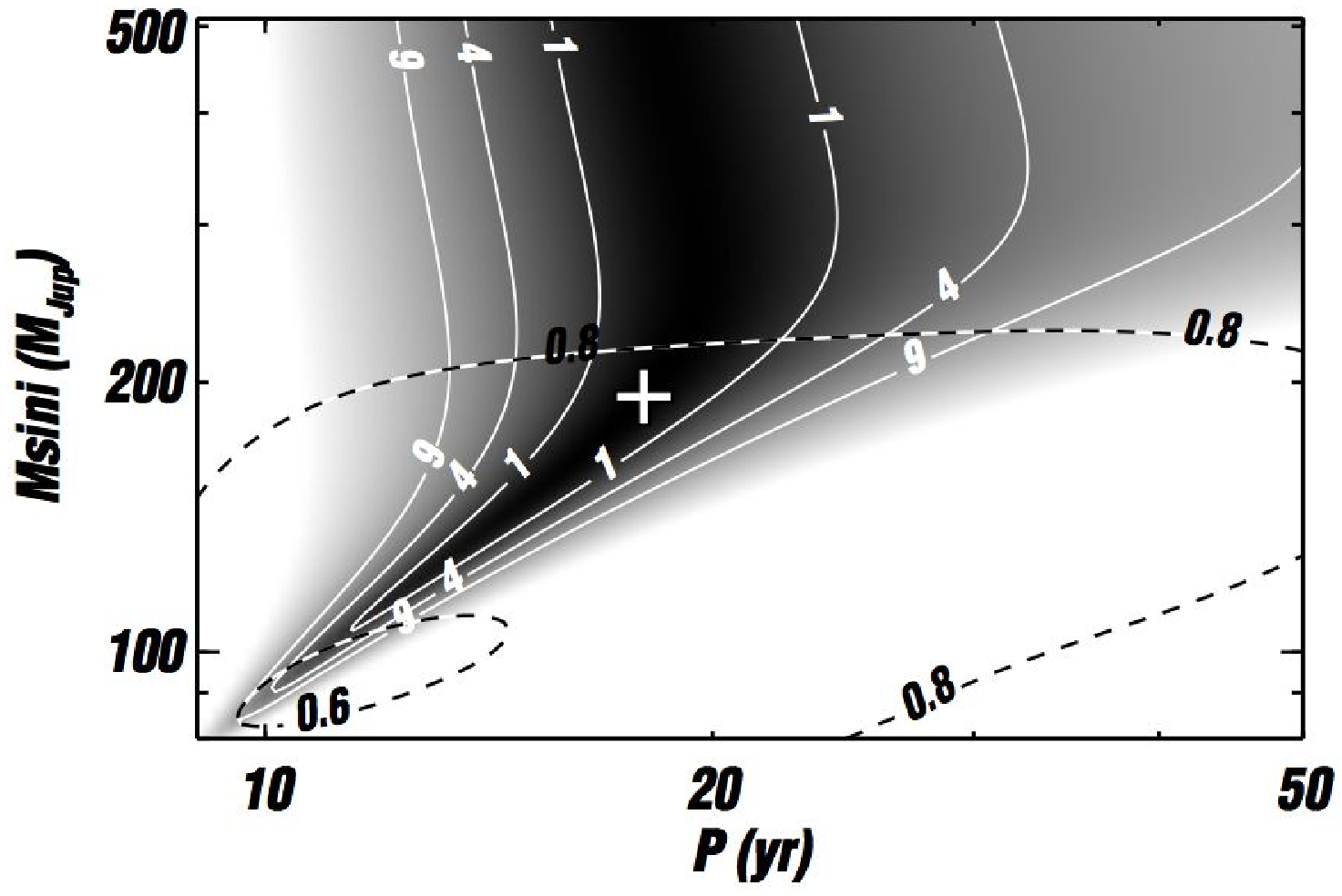}{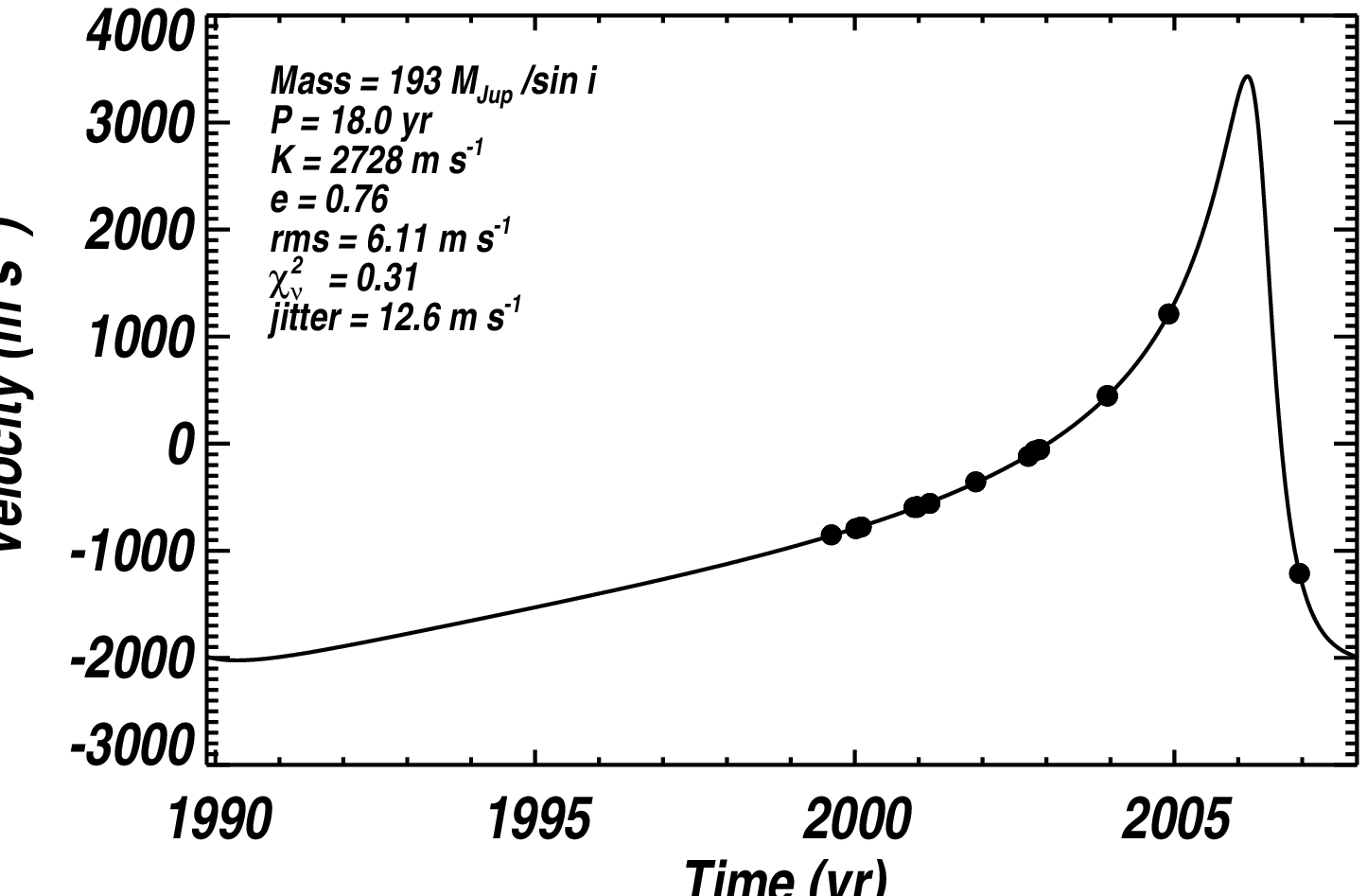}
\caption{ {\em Left}: Contours of $\chi_\nu^2$ in $M$ sin $i$ -- $P$ space for HD 29461.  The solid contour lines represent increases from the best-fit ($\chi_{\nu, {\rm min}}^2$) of 1, 4, and 9.  The location of the best-fit on the $M$ sin $i$ -- $P$ grid is indicated with a cross at 193 $M_{\rm Jup}$ and 18.0 yr, and its corresponding Keplerian orbit is plotted on the right.  Contours of eccentricity are shown as dashed lines.  The $\chi_\nu^2$ = $\chi_{\nu, {\rm min}}^2 + 4 $ contour and $e < 0.8$ constraint limits the minimum mass and period to: $91.2 < M$ sin $i$ $(M_{\rm Jup}) < 224$, $10.2 < P$ (yr) $< 25.9$.  {\em Right}: Velocity vs. time for HD 29461 (dots).  The solid line is the Keplerian orbit for the best-fit ($\chi_{\nu, {\rm min}}^2$) from the $M$ sin $i$ -- $P$ grid.} \label{plot29461}
\end{figure}

\begin{figure}
\epsscale{1}
\plottwo{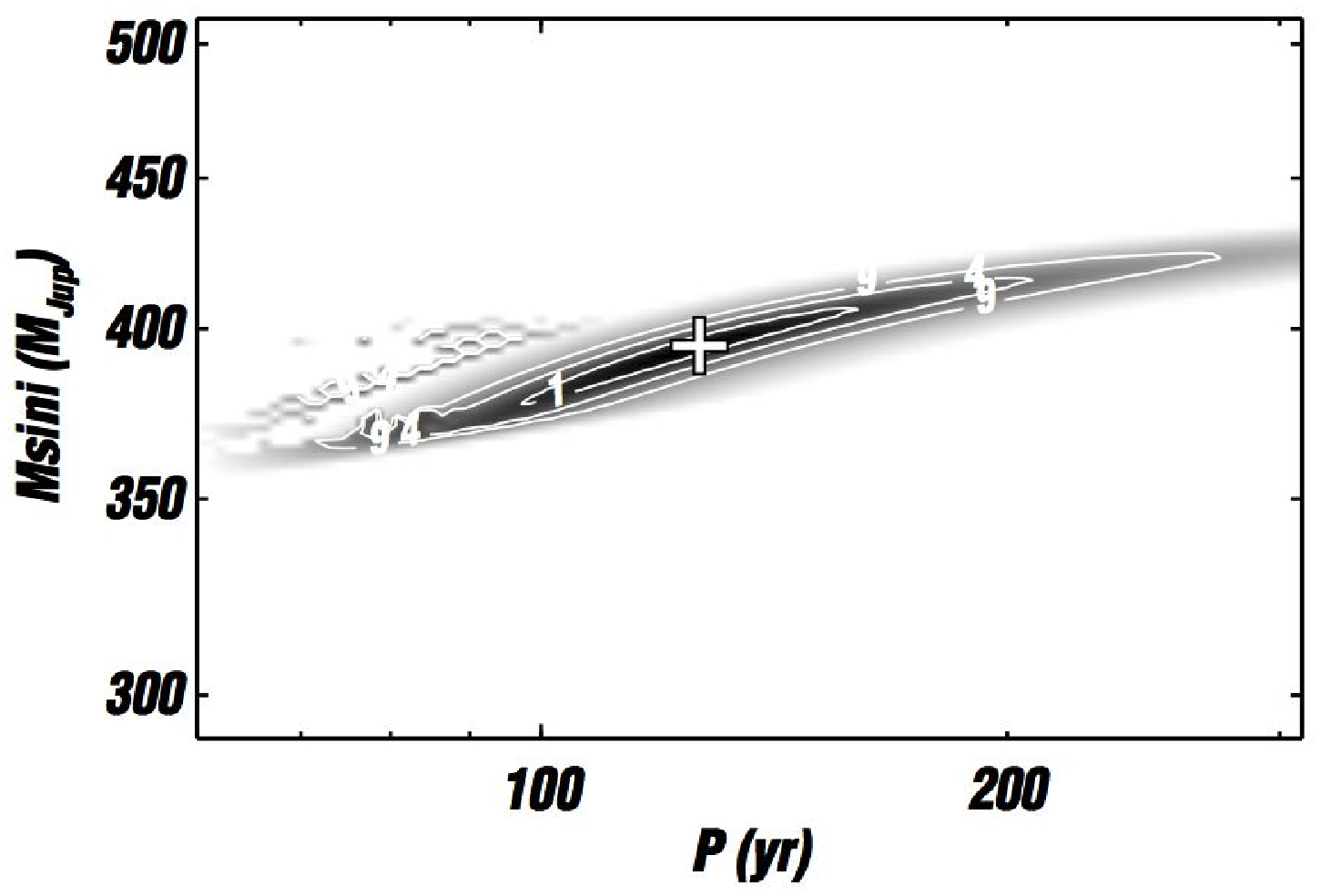}{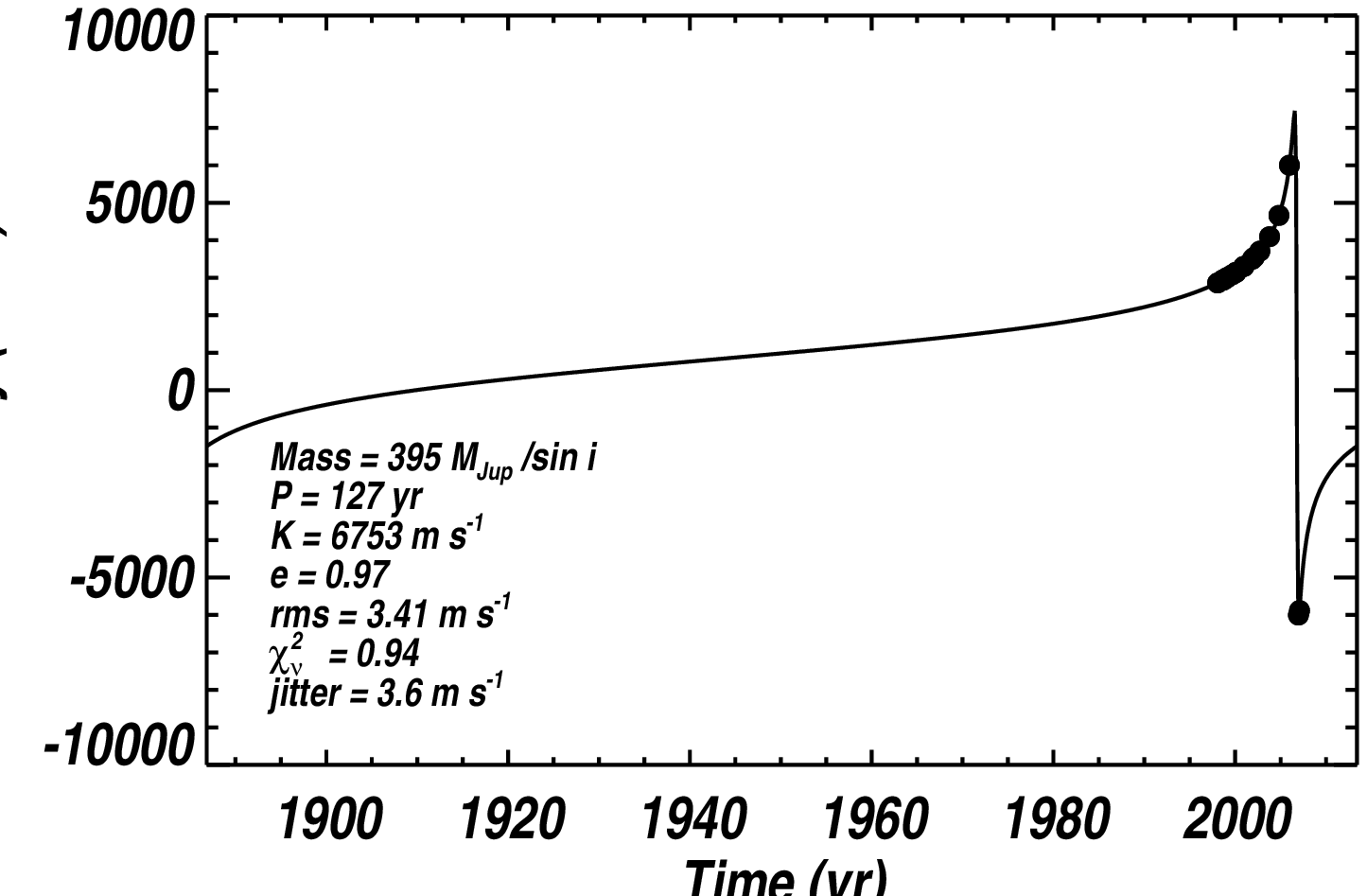}
\caption{ {\em Left}: Contours of $\chi_\nu^2$ in $M$ sin $i$ -- $P$ space for HD 31412.  The solid contour lines represent increases from the best-fit ($\chi_{\nu, {\rm min}}^2$) of 1, 4, and 9.  The location of the best-fit on the $M$ sin $i$ -- $P$ grid is indicated with a cross at 395 $M_{\rm Jup}$ and 127 yr, and its corresponding Keplerian orbit is plotted on the right.  The $\chi_\nu^2$ = $\chi_{\nu, {\rm min}}^2 + 4 $ contour limits the minimum mass and period to: $368 < M$ sin $i$ $(M_{\rm Jup}) < 415$, $75.7 < P$ (yr) $< 205$.  {\em Right}: Velocity vs. time for HD 31412 (dots).  The solid line is the Keplerian orbit for the best-fit ($\chi_{\nu, {\rm min}}^2$) from the $M$ sin $i$ -- $P$ grid.} \label{plot31412}
\end{figure}

\begin{figure}
\epsscale{1}
\plottwo{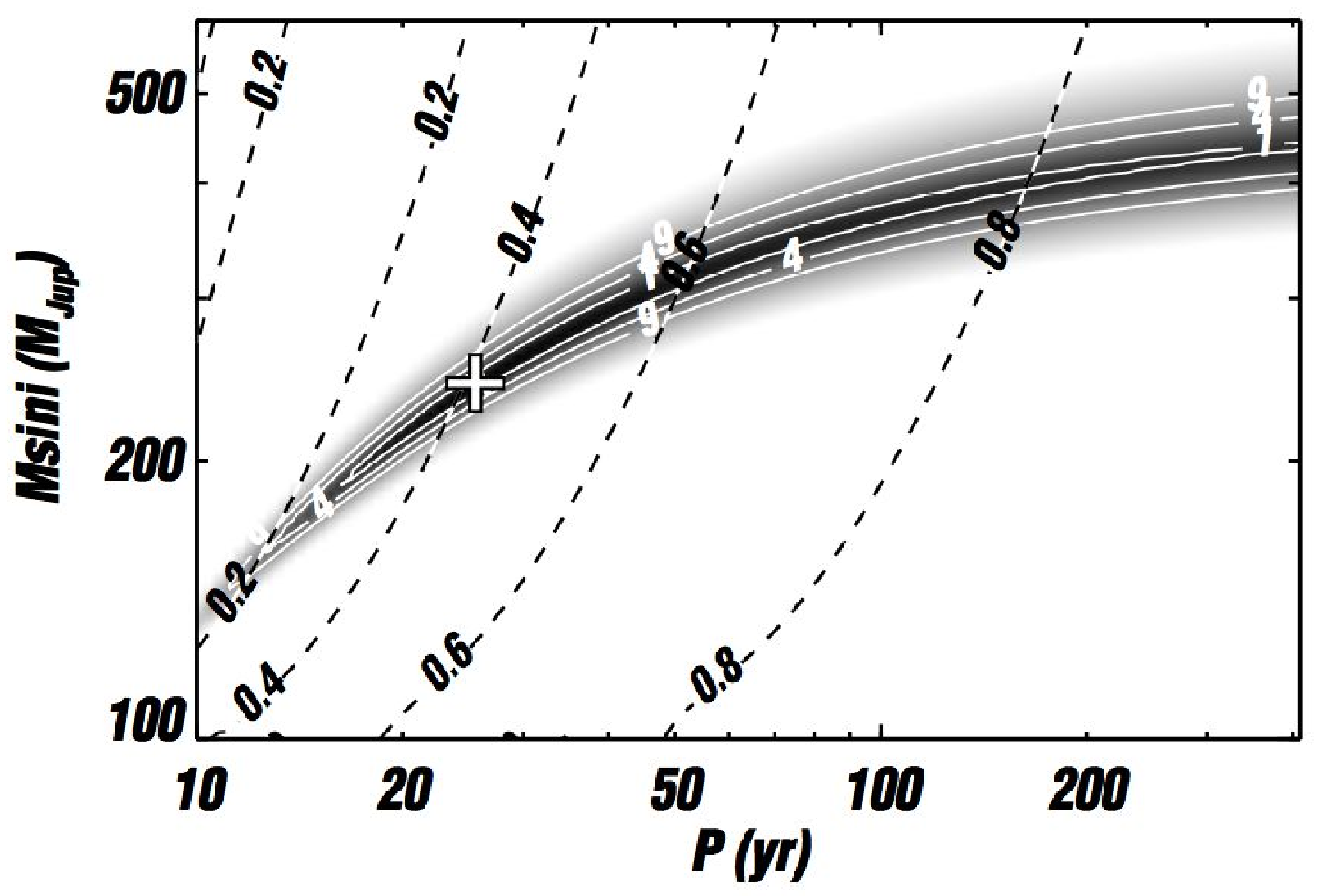}{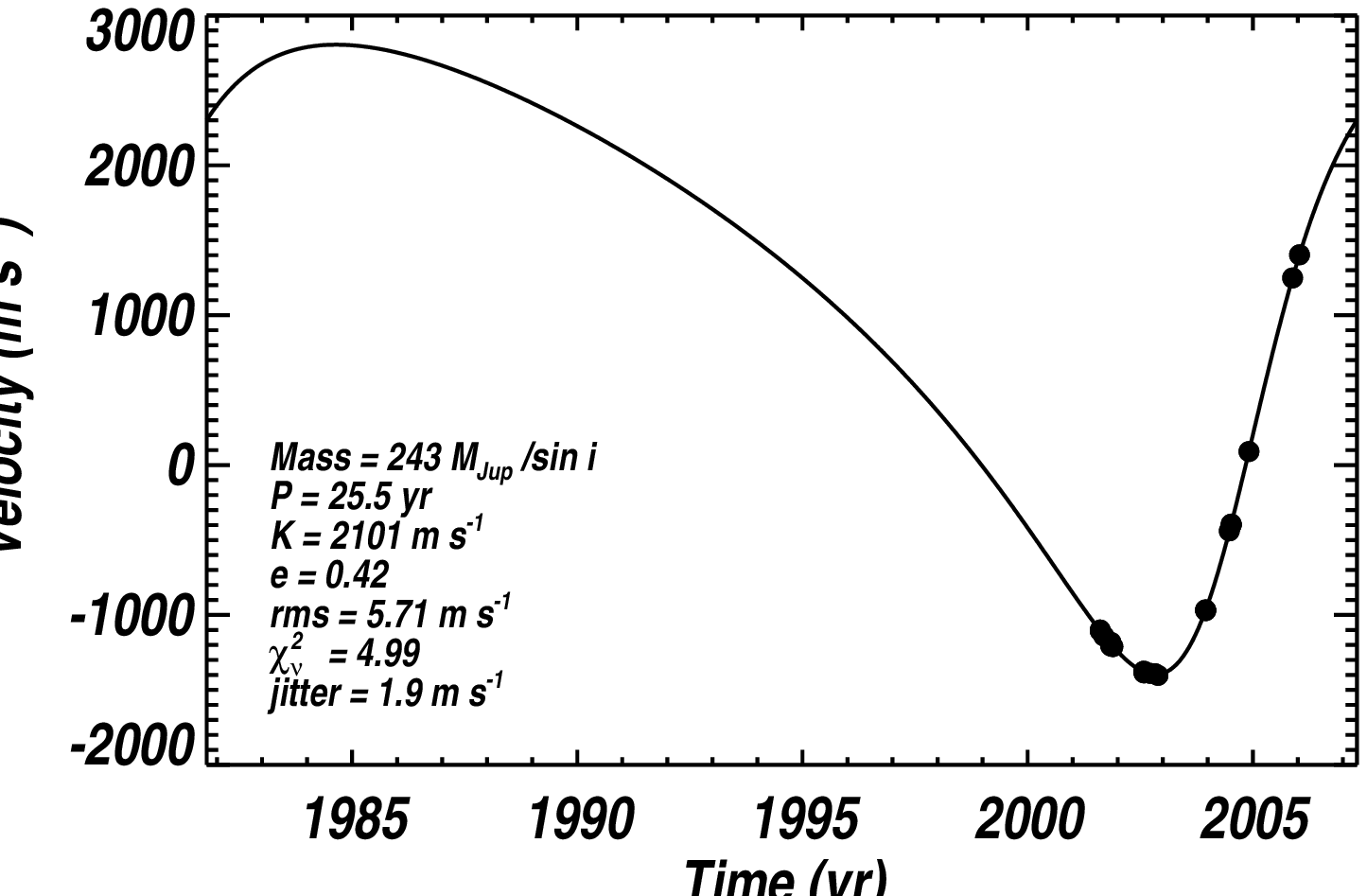}
\caption{ {\em Left}: Contours of $\chi_\nu^2$ in $M$ sin $i$ -- $P$ space for HD 5470.  The solid contour lines represent increases from the best-fit ($\chi_{\nu, {\rm min}}^2$) of 1, 4, and 9.  The location of the best-fit on the $M$ sin $i$ -- $P$ grid is indicated with a cross at 243 $M_{\rm Jup}$ and 25.5 yr, and its corresponding Keplerian orbit is plotted on the right.  Contours of eccentricity are shown as dashed lines.  The $\chi_\nu^2$ = $\chi_{\nu, {\rm min}}^2 + 4 $ contour and $e < 0.8$ constraint limits the minimum mass and period to: $163 < M$ sin $i$ $(M_{\rm Jup}) < 425$, $13.0 < P$ (yr) $< 167$.  {\em Right}: Velocity vs. time for HD 5470 (dots).  The solid line is the Keplerian orbit for the best-fit ($\chi_{\nu, {\rm min}}^2$) from the $M$ sin $i$ -- $P$ grid.} \label{plot5470}
\end{figure}

\begin{figure}
\epsscale{1}
\plottwo{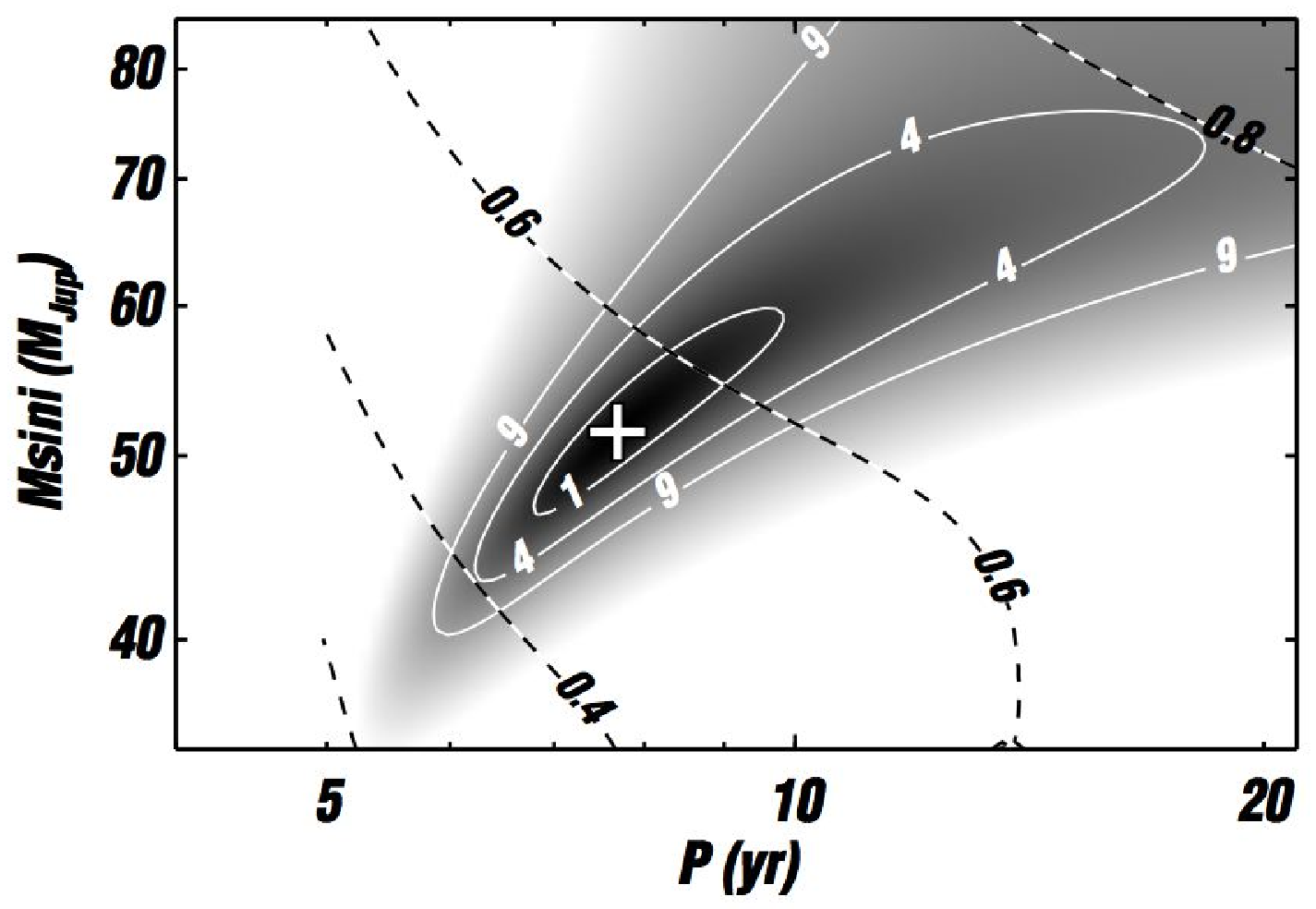}{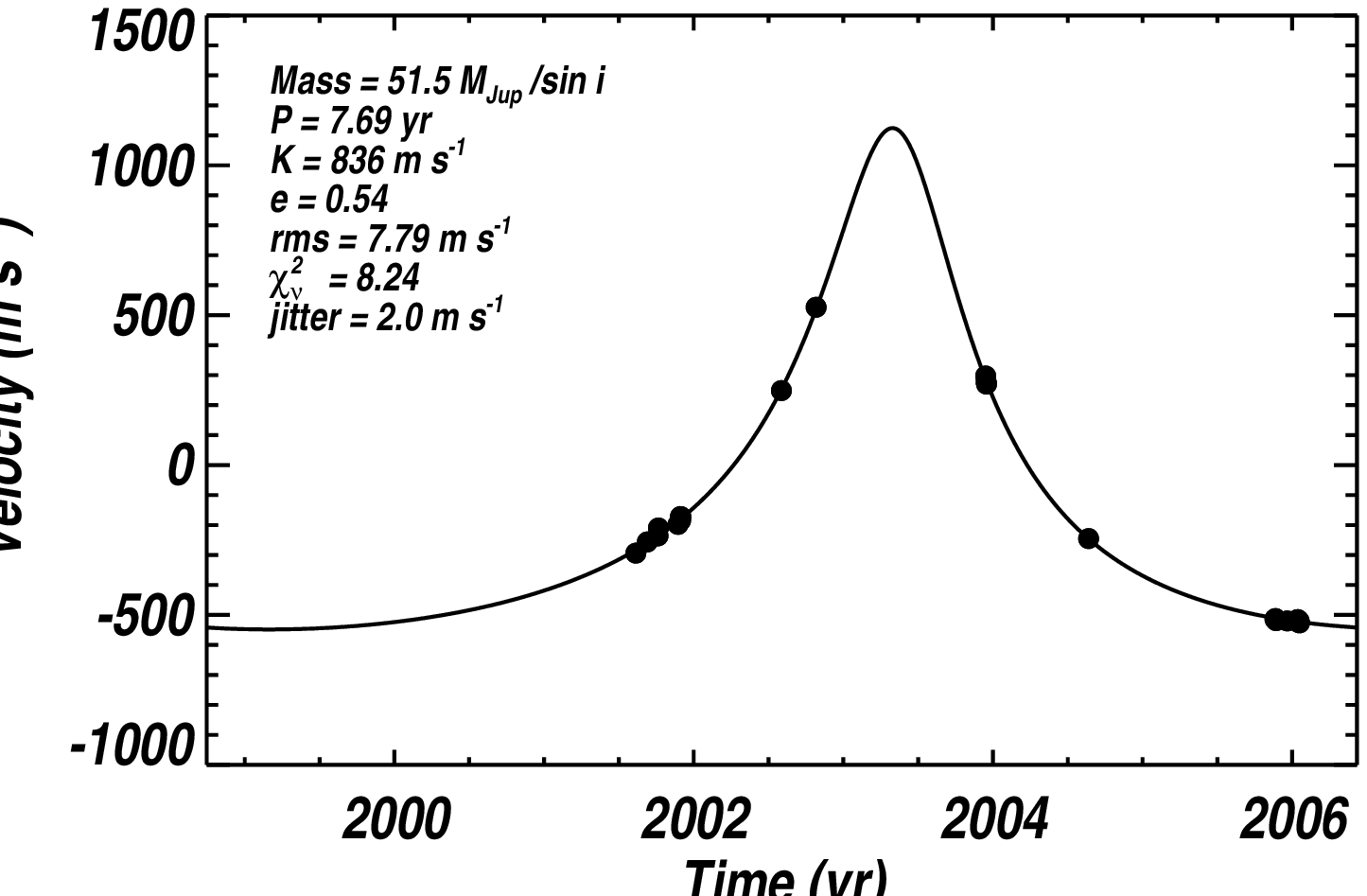}
\caption{ {\em Left}: Contours of $\chi_\nu^2$ in $M$ sin $i$ -- $P$ space for HD 8765.  The solid contour lines represent increases from the best-fit ($\chi_{\nu, {\rm min}}^2$) of 1, 4, and 9.  The location of the best-fit on the $M$ sin $i$ -- $P$ grid is indicated with a cross at 51.5 $M_{\rm Jup}$ and 7.69 yr, and its corresponding Keplerian orbit is plotted on the right.  Contours of eccentricity are shown as dashed lines.  The $\chi_\nu^2$ = $\chi_{\nu, {\rm min}}^2 + 4 $ contour and $e < 0.8$ constraint limits the minimum mass and period to: $43.0 < M$ sin $i$ $(M_{\rm Jup}) < 75.7$, $6.3 < P$ (yr) $< 18.1$.  {\em Right}: Velocity vs. time for HD 8765 (dots).  The solid line is the Keplerian orbit for the best-fit ($\chi_{\nu, {\rm min}}^2$) from the $M$ sin $i$ -- $P$ grid.} \label{plot8765}
\end{figure}

\begin{figure}
\epsscale{1}
\plottwo{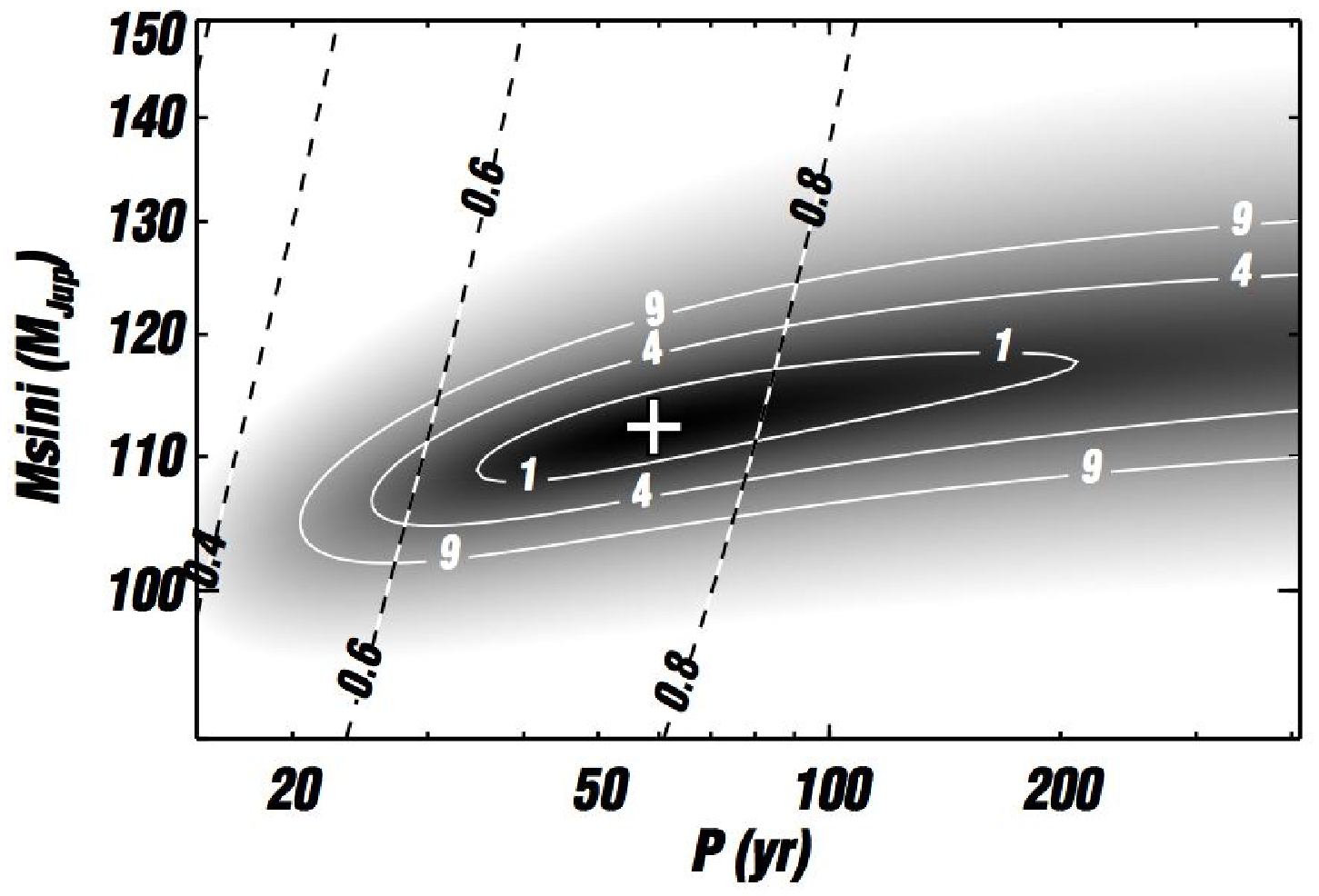}{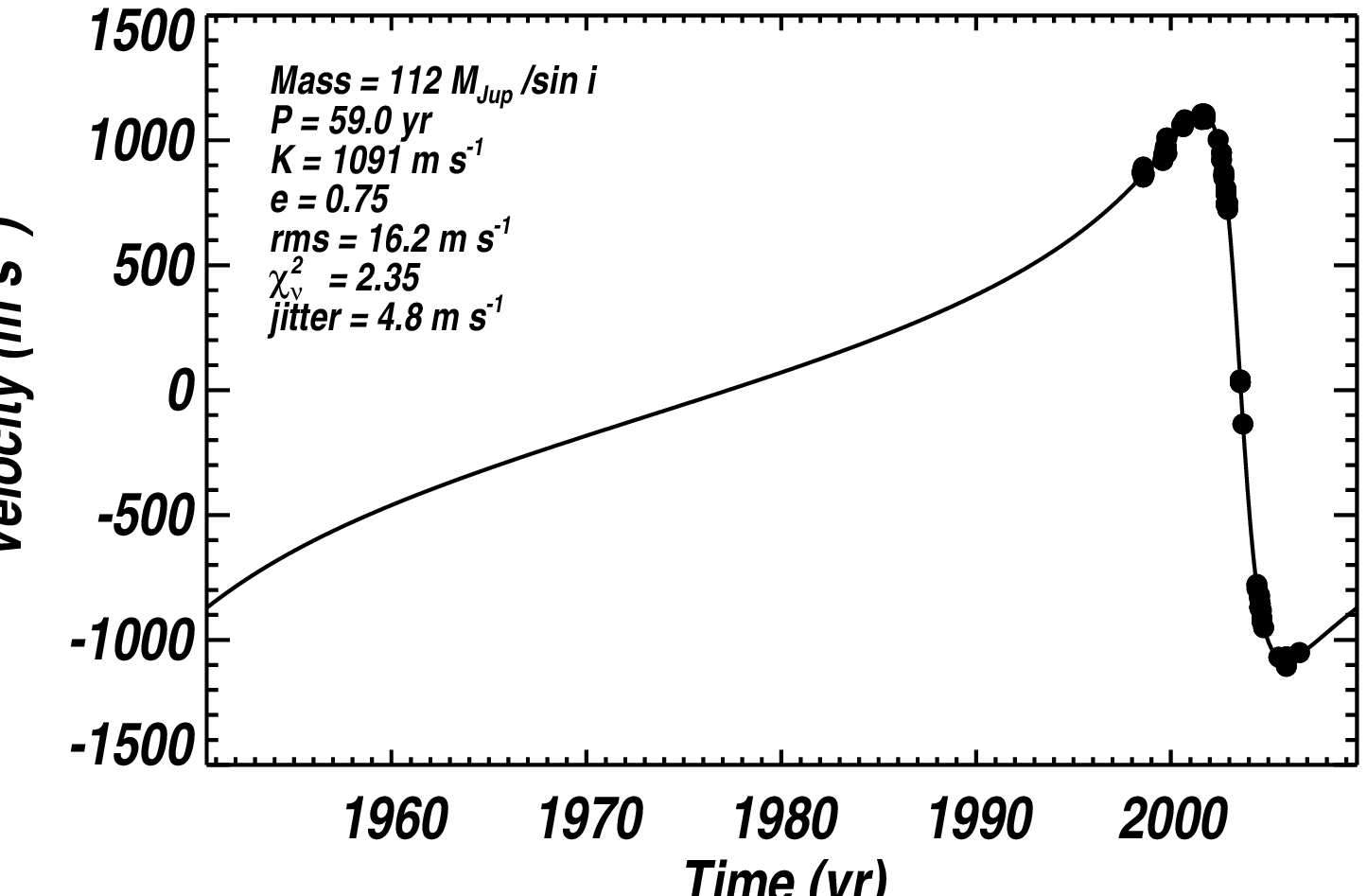}
\caption{ {\em Left}: Contours of $\chi_\nu^2$ in $M$ sin $i$ -- $P$ space for HD 199598.  The solid contour lines represent increases from the best-fit ($\chi_{\nu, {\rm min}}^2$) of 1, 4, and 9.  The location of the best-fit on the $M$ sin $i$ -- $P$ grid is indicated with a cross at 112 $M_{\rm Jup}$ and 59.0 yr, and its corresponding Keplerian orbit is plotted on the right.  Contours of eccentricity are shown as dashed lines.  The $\chi_\nu^2$ = $\chi_{\nu, {\rm min}}^2 + 4 $ contour and $e < 0.8$ constraint limits the minimum mass and period to: $105 < M$ sin $i$ $(M_{\rm Jup}) < 120$, $25.6 < P$ (yr) $< 85.3$.  {\em Right}: Velocity vs. time for HD 199598 (dots).  The solid line is the Keplerian orbit for the best-fit ($\chi_{\nu, {\rm min}}^2$) from the $M$ sin $i$ -- $P$ grid.} \label{plothd199598}
\end{figure}

\begin{figure}
\epsscale{1}
\plottwo{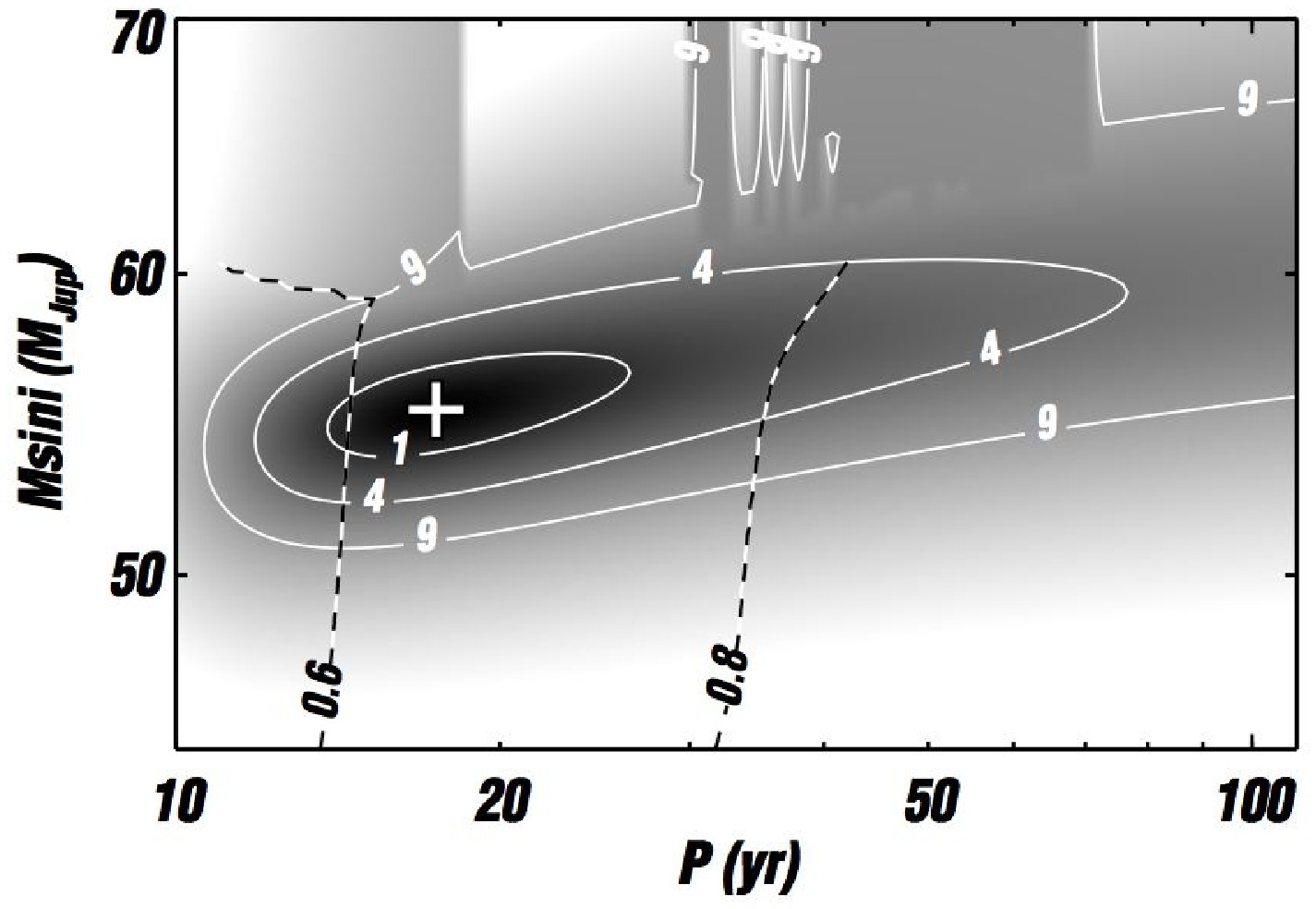}{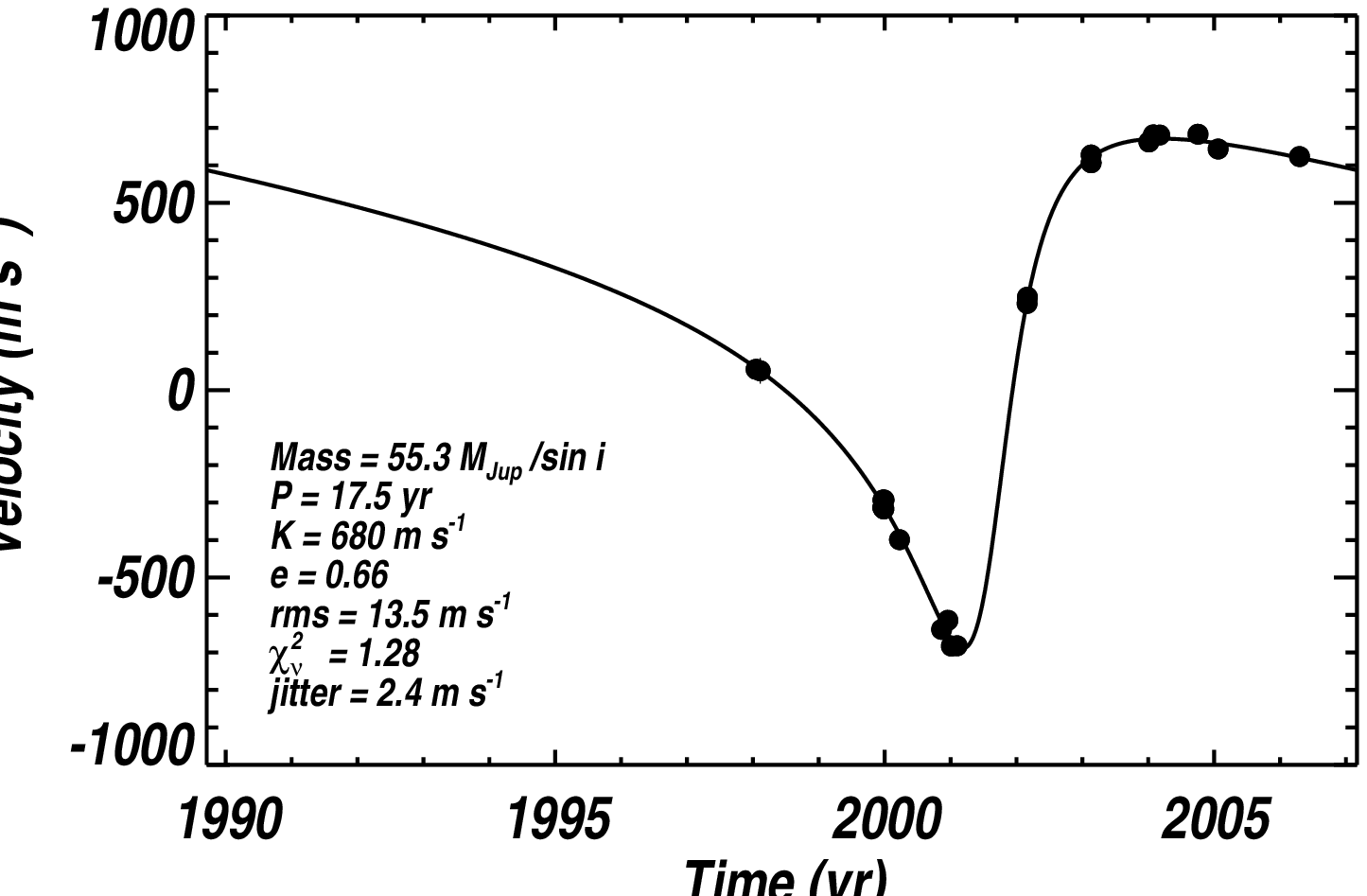}
\caption{ {\em Left}: Contours of $\chi_\nu^2$ in $M$ sin $i$ -- $P$ space for HD 72780.  The solid contour lines represent increases from the best-fit ($\chi_{\nu, {\rm min}}^2$) of 1, 4, and 9.  The location of the best-fit on the $M$ sin $i$ -- $P$ grid is indicated with a cross at 55.3 $M_{\rm Jup}$ and 17.5 yr, and its corresponding Keplerian orbit is plotted on the right.  Contours of eccentricity are shown as dashed lines.  The $\chi_\nu^2$ = $\chi_{\nu, {\rm min}}^2 + 4 $ contour and $e < 0.8$ constraint limits the minimum mass and period to: $52.4 < M$ sin $i$ $(M_{\rm Jup}) < 60.4$, $11.8 < P$ (yr) $< 41.7$.  {\em Right}: Velocity vs. time for HD 72780 (dots).  The solid line is the Keplerian orbit for the best-fit ($\chi_{\nu, {\rm min}}^2$) from the $M$ sin $i$ -- $P$ grid.} \label{plothd72780}
\end{figure}

\begin{figure}
\epsscale{1}
\plottwo{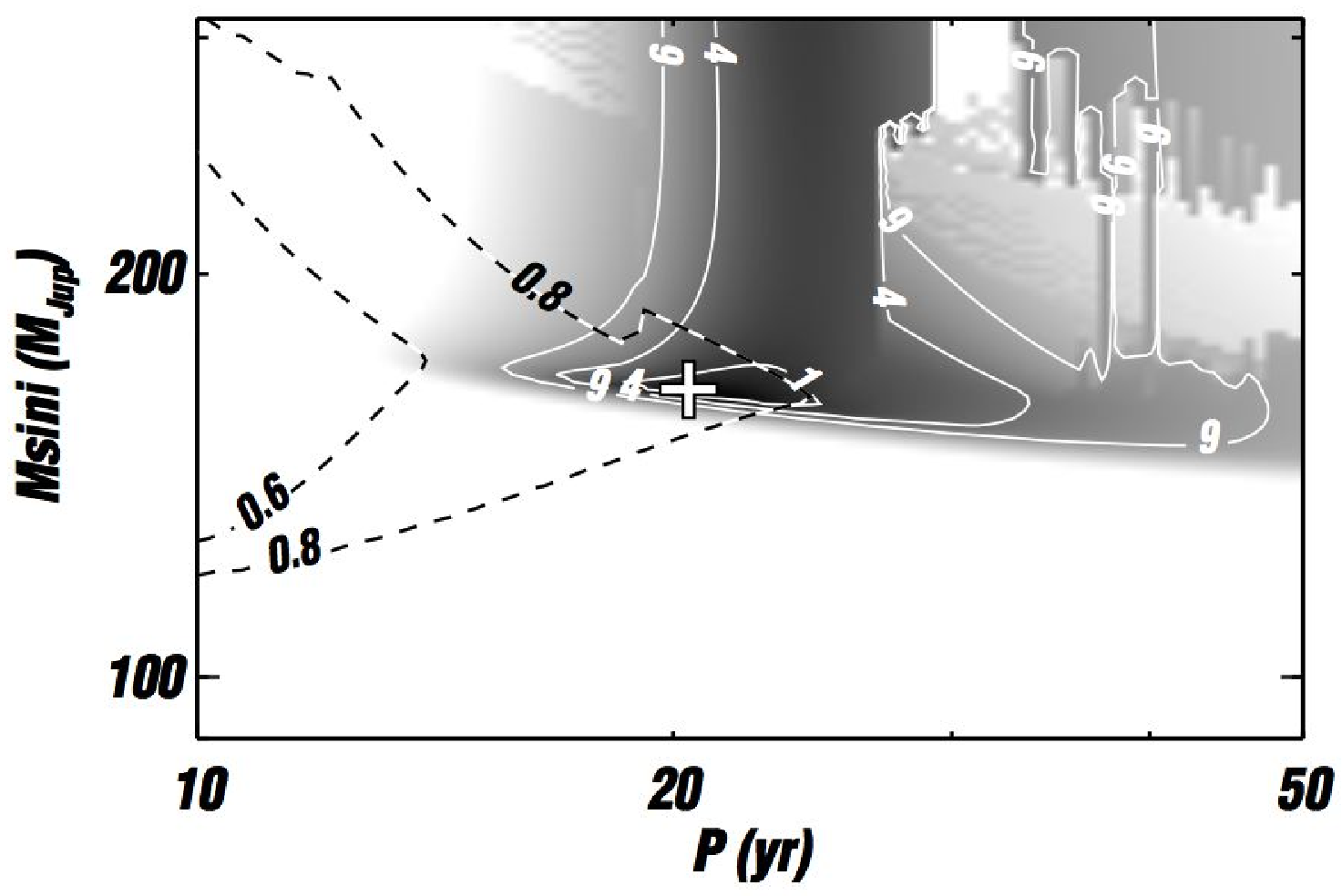}{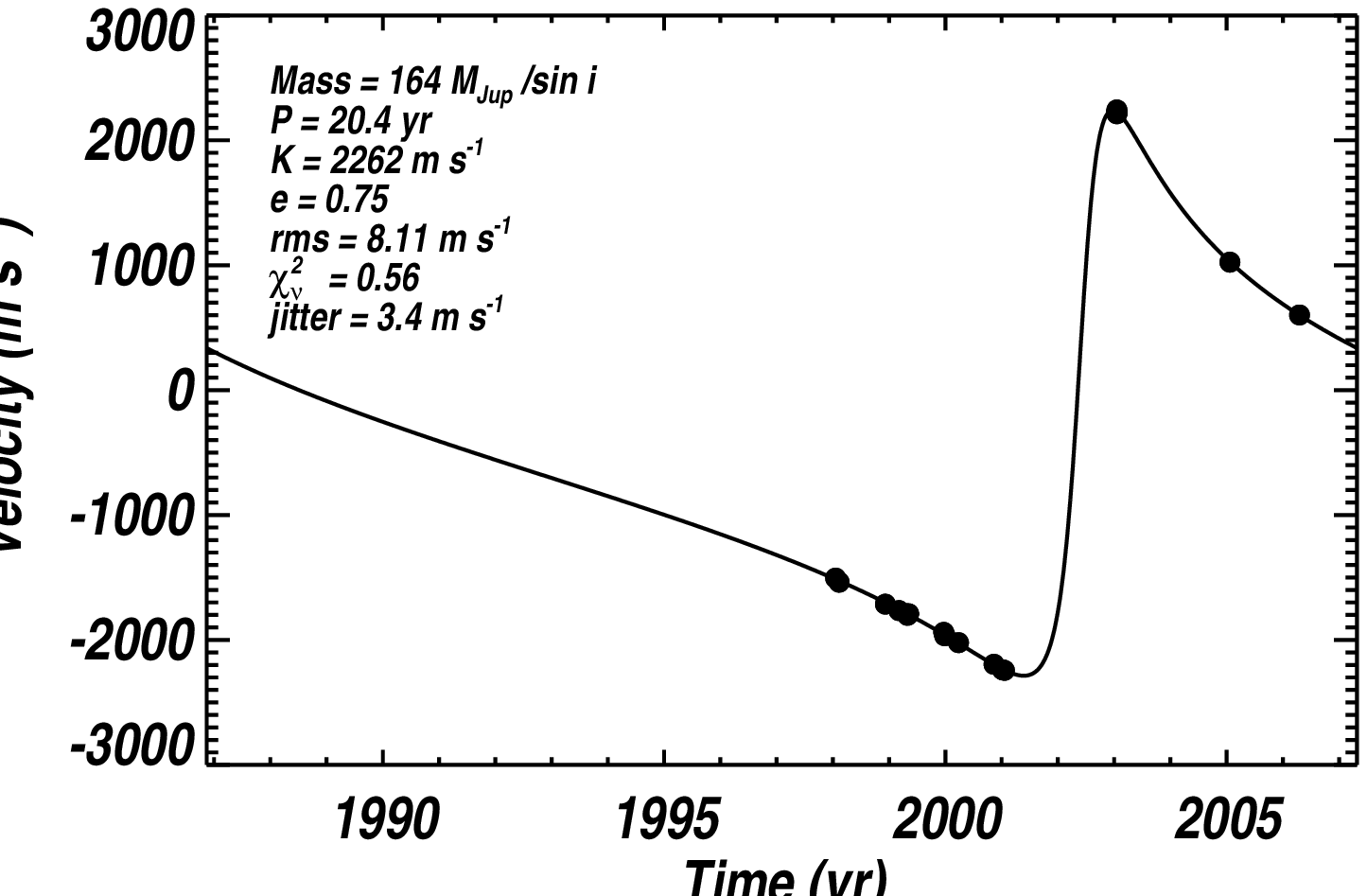}
\caption{ {\em Left}: Contours of $\chi_\nu^2$ in $M$ sin $i$ -- $P$ space for HD 73668.  The solid contour lines represent increases from the best-fit ($\chi_{\nu, {\rm min}}^2$) of 1, 4, and 9.  The location of the best-fit on the $M$ sin $i$ -- $P$ grid is indicated with a cross at 164 $M_{\rm Jup}$ and 20.4 yr, and its corresponding Keplerian orbit is plotted on the right.  Contours of eccentricity are shown as dashed lines.  The $\chi_\nu^2$ = $\chi_{\nu, {\rm min}}^2 + 4 $ contour and $e < 0.8$ constraint limits the minimum mass and period to: $160 < M$ sin $i$ $(M_{\rm Jup}) < 181$, $17.1 < P$ (yr) $< 24.1$.  {\em Right}: Velocity vs. time for HD 73668 (dots).  The solid line is the Keplerian orbit for the best-fit ($\chi_{\nu, {\rm min}}^2$) from the $M$ sin $i$ -- $P$ grid.} \label{plothd73668}
\end{figure}

\begin{figure}
\epsscale{1}
\plottwo{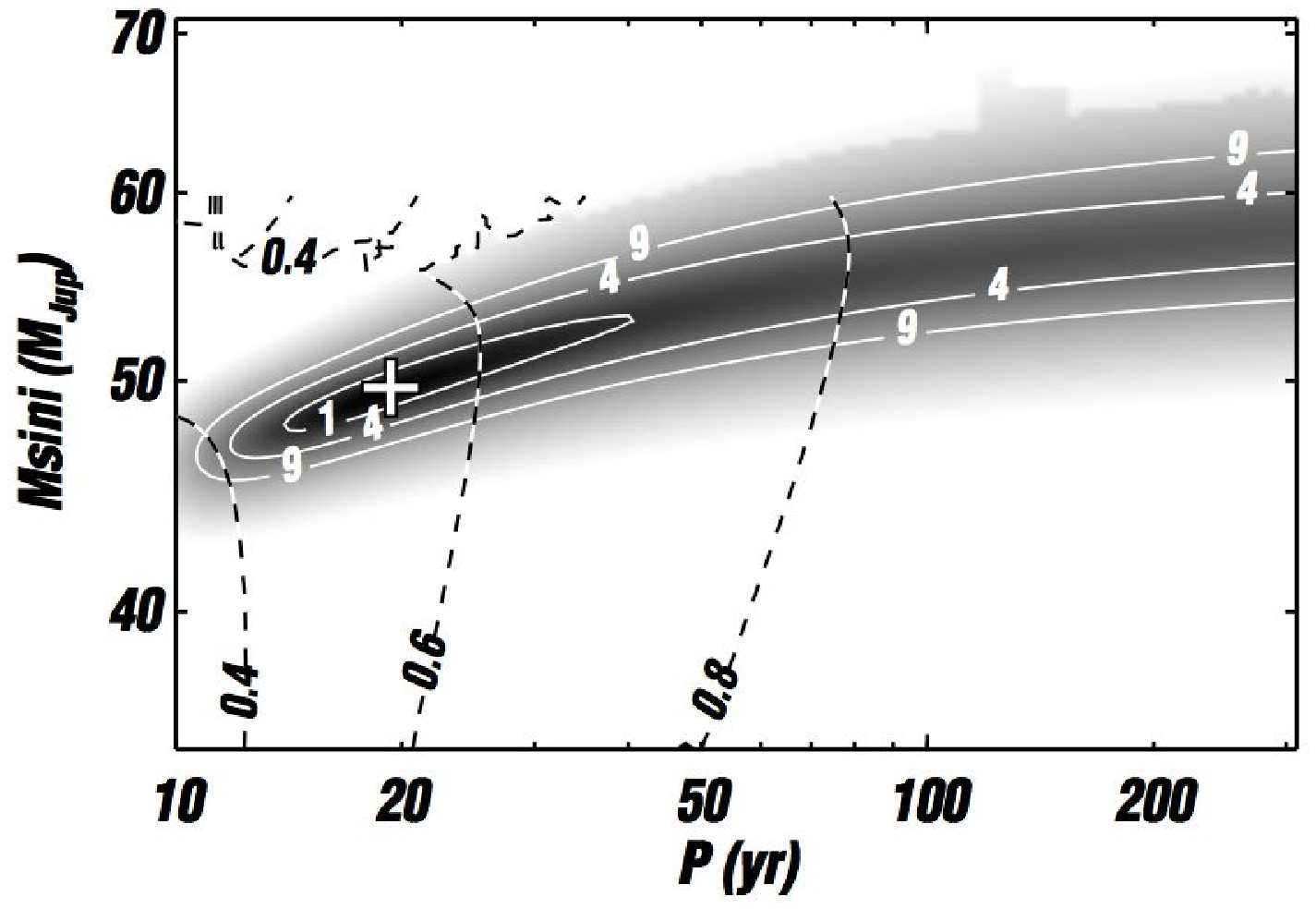}{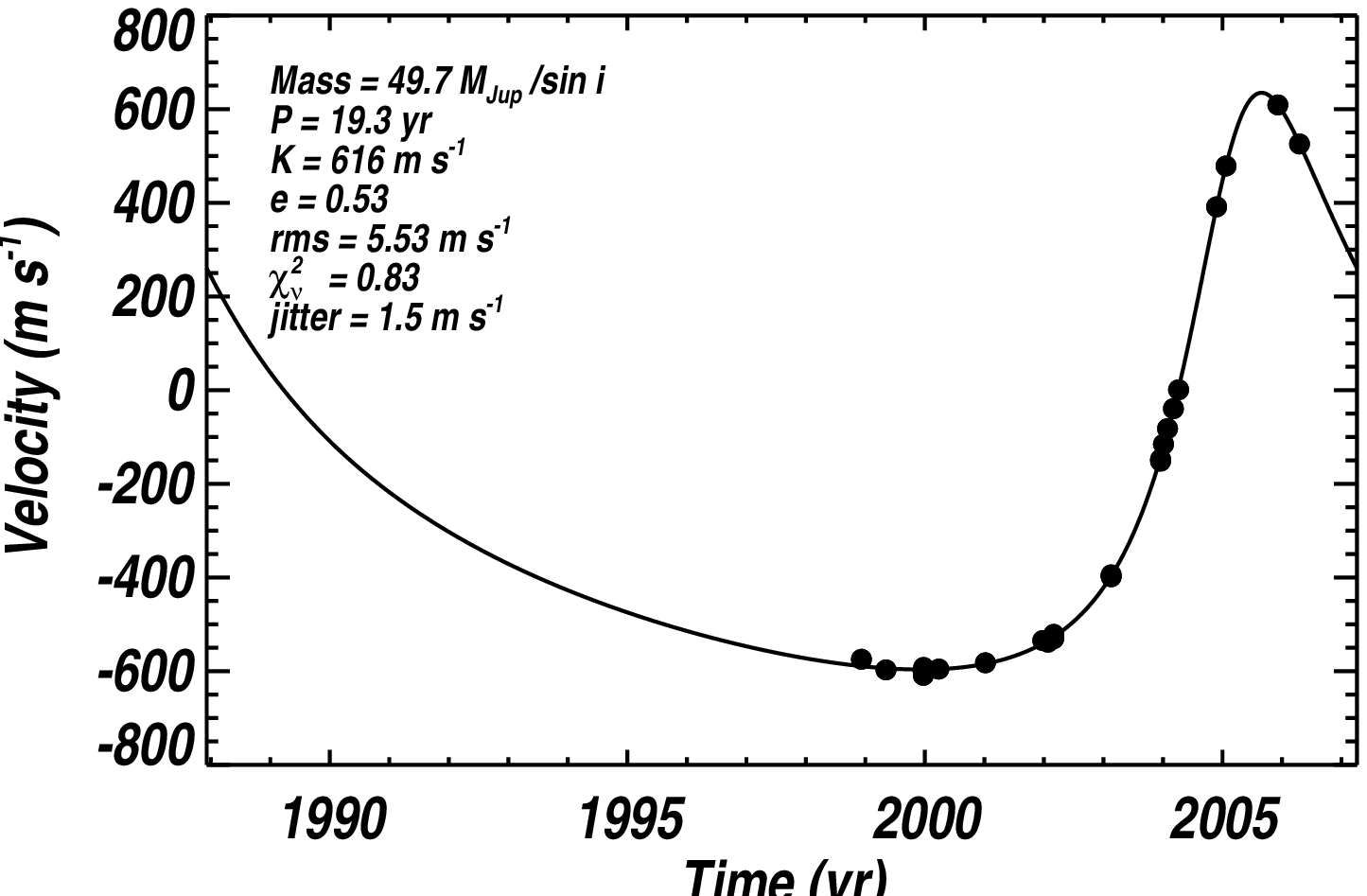}
\caption{ {\em Left}: Contours of $\chi_\nu^2$ in $M$ sin $i$ -- $P$ space for HD 74014.  The solid contour lines represent increases from the best-fit ($\chi_{\nu, {\rm min}}^2$) of 1, 4, and 9.  The location of the best-fit on the $M$ sin $i$ -- $P$ grid is indicated with a cross at 49.7 $M_{\rm Jup}$ and 19.3 yr, and its corresponding Keplerian orbit is plotted on the right.  Contours of eccentricity are shown as dashed lines.  The $\chi_\nu^2$ = $\chi_{\nu, {\rm min}}^2 + 4 $ contour and $e < 0.8$ constraint limits the minimum mass and period to: $46.6 < M$ sin $i$ $(M_{\rm Jup}) < 57.3$, $11.9 < P$ (yr) $< 77.4$.  {\em Right}: Velocity vs. time for HD 74014 (dots).  The solid line is the Keplerian orbit for the best-fit ($\chi_{\nu, {\rm min}}^2$) from the $M$ sin $i$ -- $P$ grid.} \label{plothd74014}
\end{figure}

An initial grid with a large range in period and minimum mass was constructed to determine the region of \msini\ -- $P$ space to iterate on.  In Figures \ref{plot142229} -- \ref{plothd74014} we show the more focused grids of \rchi\ in \msini\ --  $P$ space, with contour levels of 1, 4, and 9 from the minimum ($\chi_{\nu,{\rm min}}^2$) corresponding to 1, 2, and 3-sigma confidence levels.  One hundred logarithmic steps in \msini\ and $P$ are used for each focused grid with step sizes dependent on the range displayed on the axes in Figures 2--14.  Contours of eccentricity are shown in these figures in increments of 0.2.  Also presented in the figures are the radial velocity measurements for each star along with the Keplerian orbit corresponding to the best-fit on the \msini\ -- $P$ grid (indicated by a cross).  The best-fit \rchi\ quoted in these figures includes the contribution from the estimated photospheric jitter, however we caution the reader to note the uncertainty in this value due to the uncertain nature of the jitter.  The \msini\ -- $P$ grids were run without the jitters for this reason and because the results are insensitive to the jitter levels assumed at these large amplitudes.  As a test, we find that when including jitter, the minimum \rchi\ in the \msini -- $P$ grid corresponds to the best-fit solution we obtain when excluding jitter.  The jitters given in the figures were derived from recent improvements to the jitter algorithm of \citet{wright05}.  We note that while it is often customary to remove linear velocity trends before performing the Keplerian fits, we refrain from doing so here because the limited phase coverage prevents a well-determined linear solution.

Using the $\chi_{\nu}^2 = \chi_{\nu,{\rm min}}^2 + 4$ contour as a constraint, we can place limits on the period and minimum mass.  We also implement an additional eccentricity constraint by noting that for those spectroscopic binaries with $P > 10$ yr in the Ninth Catalogue of Spectroscopic Binary Orbits \citep{pourbaix04}, approximately 90\% have an eccentricity \elimit.  Therefore, we will assume the companions reported here have \elimit\ unless their eccentricity explicitly exceeds this limit.  We report the possible range in minimum mass and period along with the semimajor axis, $a$, in Table \ref{limits}.  Note that upper limits are absent when the $\chi_{\nu}^2 = \chi_{\nu,{\rm min}}^2 + 4$ contour extends off of the grid and the eccentricity constraint is unable to provide additional limits.

\subsection{New Companions} \label{companions}

The orbit for the companion to HD 142229 (Figure \ref{plot142229}) is not well constrained.  The location of the best-fit on the \msini\ -- $P$ grid is on its upper edge (arrow), indicating that higher masses would lead to acceptable fits.  However, a higher mass is unlikely given that a companion of mass greater than 1000 \mjup\ ($\sim$ 1 \msun) would probably have been detected either visually or spectroscopically.

While the period for the companion to HD 150554 (Figure \ref{plot150554}) is uncertain, the minimum mass is well constrained due to phase coverage near the turning points.  If we assume \elimit\ then the minimum mass for the companion is bound to the brown dwarf mass range with a period between 8 and 34 yr.

HD 211681 (Figure \ref{plot211681}) is another example of how instituting an eccentricity limit of \elimit\ can help constrain the period.  The companion has a period of approximately between 10 and 100 yr.  Though the upper limit of \msini\ is 102 \mjup, the lower limit does not exclude the companion from the brown dwarf mass regime.

Only lower limits to \msini\ and $P$ were obtained for the companion to HD 215578 (Figure \ref{plot215578}) because the $\chi_{\nu}^2 = \chi_{\nu,{\rm min}}^2 + 4$ contour extended off of the grid.  With a lower limit for \msini\ of $\sim 0.5$ \msun, this companion is one the largest of those reported in this paper.

Lower limits to \msini\ and $P$ are also reported for HD 217165 (Figure \ref{plot217165}) for the same reason as HD 215578.  Given the curvature present in the radial velocity data, a wide range of minimum masses yielded acceptable fits.  While the best-fit is $\sim$ 200 \mjup, the lower limit for \msini\ extends into the brown dwarf mass range.  Note that the break in the \rchi\ surface between 10 and 20 yr is an artifact of proceeding to the next pixel in the grid using the previous pixel's solution.  Traversing the grid in this fashion allows us to minimize random departures from the best Keplerian solution.

The most recent radial velocity observation for HD 29461 (Figure \ref{plot29461}) was able to put strong constraints on the minimum mass and period for the companion.  Assuming \elimit, the object is most likely a low-mass star with a lower limit for \msini\ of $\sim$ 90 \mjup.  The period is between $\sim$ 10 and 25 yr.

The eccentricity criteria of \elimit\ was not implemented for HD 31412 (Figure \ref{plot31412}) given its highly eccentric orbit.  The companion has a minimum mass between $\sim$ 370 and 415 \mjup, and period between $\sim$ 75 and 200 yr.

The companion to HD 5470 (Figure \ref{plot5470}) is most likely a low-mass star.  Assuming \elimit, the lower limit for \msini\ is $\sim$ 160 \mjup\ and the upper limit $\sim$ 400 \mjup.  The period is as short as $\sim$ 13 yr and as long as $\sim$ 170 yr, putting the companion as far away as 35 AU from its host star.

The $\chi_{\nu}^2 = \chi_{\nu,{\rm min}}^2 + 4$ contour limits the companion for HD 8765 (Figure \ref{plot8765}) to the brown dwarf mass regime with an upper limit for \msini\ of $\sim$ 76 \mjup.  The period ranges between $\sim$ 6 and $\sim$ 18 yr.  We note that HD 8765 is given an acceleration solution (G) in the {\em Hipparcos} catalogue indicating that a quadratic or cubic astrometric solution was obtained using the position, proper motion, and parallax \citep{lindegren97}.

The minimum mass for the companion to HD 199598 (Figure \ref{plothd199598}) is well-determined due to radial velocity measurements around the turning points, providing strong constraints on the semiamplitude.  Assuming \elimit, the companion is a low-mass star with minimum mass between 105 and 120 \mjup\ and period between 25 and 85 yr.

The companion to HD 72780 (Figure \ref{plothd72780}) is another potential brown dwarf candidate with a minimum mass constrained between $\sim$ 52 and 60 \mjup, assuming \elimit.  The period is less constrained, ranging between $\sim$ 12 and 42 yr.

The low-mass star that orbits HD 73668 (Figure \ref{plothd73668}) is in a fairly eccentric orbit as seen in the radial velocity measurements.  Assuming \elimit, the minimum mass and period are very well constrained.  The minimum mass ranges between $\sim$ 160 and 180 \mjup, while the period is bound between $\sim$ 17 and 24 yr.  We note that HD 73668 is given a component solution (C) in the {\em Hipparcos} catalogue indicating that a linear astrometric solution was obtained for it and another nearby visible component.

Phase coverage near the turning points of HD 74014 (Figure \ref{plothd74014}) puts strong limits on its semiamplitude and in turn the minimum mass of the companion.  Assuming \elimit, the the minimum mass is between $\sim$ 47 and 57 \mjup.  The period is less constrained, with a lower limit of $\sim$ 12 yr and an upper limit of $\sim$ 77 yr.

\section{Discussion} \label{disc}

Of the 14 new companions to stars in the CCPS reported here, 7 have \msini\ extending into the brown dwarf mass regime.  Of those 7, 5 have upper limits for \msini\ below the brown dwarf -- stellar mass boundary (HD 150554, HD 167665, HD 8765, HD 72780, HD 74014), making them potential brown dwarf candidates.  The sin $i$ degeneracy ultimately determines the true mass of these companions.

{\em Hipparcos} data can serve as a useful tool in constraining orbital inclination and thus the companion mass.  There are two ways in which one typically goes about doing this.  The first method makes use of flags in the {\em Hipparcos} Double and Multiple Systems Annex indicating an astrometric perturbation.  One then assumes a lower limit to the astrometric perturbation detectable by {\em Hipparcos}, thus providing a lower limit for the inclination (see Equation 1 of \citet{pourbaix01}) and an upper limit for the mass of the companion.  This simple technique should serve only as a rough guide for estimating upper limits to masses given that there is no consensus on what the detectability threshold is for {\em Hipparcos}.  The second technique uses the {\em Hipparcos} Intermediate Astrometric Data along with spectroscopic data to perform a detailed 12-parameter fit to estimate the inclination \citep{reffert06}.  Both techniques only yield meaningful results when attempting to constrain inclinations for those orbits with periods less than or comparable to the {\em Hipparcos} mission lifetime ($\sim$ 4 yr).  Therefore, given the long period nature of the companions presented in this paper, we refrain from trying to present constraints on inclination with {\em Hipparcos} data.

Coronographic studies with adaptive optics (AO) can aid in placing further constraints on the masses of the brown dwarf candidates.  Limiting factors in such observations include flux contrast ratios and the separation on the sky of the brown dwarf and the host star.

Combining upper limit \msini\ estimates for our brown dwarf candidates with stellar ages from \citet{takeda07}, we can estimate near-IR apparent magnitudes using brown dwarf evolutionary models \citep{burrows97, burrows06}.  We assume the brown dwarf and the host star are coeval and at the same distance.  The host stars have $K \sim$ 5 -- 7.  Implementing filter transmission curves from 2MASS \citep{cohen03} we find the brown dwarf candidates have $K \sim$ 17 -- 21, resulting in contrasts of $\sim$ 11 -- 14 mag ($\sim 10^{-4.6} - 10^{-5.7}$).  If we assume all of the brown dwarf candidates have a mass of 73.3 \mjup, the highest mass available in the models of \citet{burrows97} for which log $g$ and \teff\ are tabulated, we obtain contrasts of $\sim 10^{-3.9} - 10^{-5.3}$ in $K$--band.  These contrast levels would make detectability with AO challenging, especially if the orbital periods and separations reside towards the lower limits derived from the \msini\ -- $P$ grid.  However, null detections would be helpful in providing upper limits on the masses of the brown dwarf candidates.

In general, the relatively older ages of stars in the CCPS implies that potential brown dwarf companions will be quite faint.  One can conclude then that imaging brown dwarfs in our survey within $\sim$ 10 AU will prove to be a difficult task due to large flux contrast ratios at short projected separations.

We have reported radial velocities for 14 new companions to stars on the CCPS target list.  For HD 167665 we were able to fit a Keplerian orbit with well-determined orbital parameters.  For the remaining stars, we explored Keplerian fits across a grid of values for \msini\ and $P$.  The technique allowed us to place constraints on the minimum masses, periods, and semimajor axes for these limited phase coverage targets.  We find that 5 of the companions reported in this work have \msini\ firmly established in the brown dwarf mass range at orbital distances of $\sim$ 4 -- 18 AU.  Imaging with adaptive optics could further aid in constraining these orbits.  Confirmation of these companions as brown dwarfs would add more examples of them to the rare sample that exists in the brown dwarf desert.

\acknowledgments

The authors wish to acknowledge the collective efforts of those involved with the CCPS including that of the staffs of both Keck and Lick Observatories.  S. G. P. thanks John Garrett and Stephen Eikenberry for their instruction and encouragement in pursuing a career in science.  S. S. V. acknowledges support from NSF grant AST-0307493.  We are appreciative of the UCO/Lick -- Keck TAC for the opportunity given to conduct this research.  This research has made use of the SIMBAD database, operated at CDS, Strasbourg, France.  The authors wish to recognize and acknowledge the very significant cultural role and reverence that the summit of Mauna Kea has always had within the indigenous Hawaiian community.  We are most fortunate to have the opportunity to conduct observations from this mountain.



\clearpage

\clearpage
\clearpage
\clearpage
\clearpage
\clearpage
\clearpage
\clearpage
\clearpage
\clearpage
\clearpage
\clearpage
\clearpage
\clearpage
\clearpage


\begin{deluxetable}{lcccccccccc}
\tabletypesize{\scriptsize}
\tablewidth{0pc}
\tablecolumns{11}
\tablecaption{Stellar Properties \label{stars}}
\tablehead{
\colhead{} & \colhead{} & \colhead{} & \colhead{$d$} & \colhead{$M_V$} & \colhead{\mstar} & \colhead{\teff} & \colhead{} & \colhead{} & \colhead{V sin $i$} & \colhead{} \\
\colhead{HD} & \colhead{HIP} & \colhead{Spectral Type} & \colhead{(pc)} & \colhead{(mag)} & \colhead{(\msun)} & \colhead{(K)} & \colhead{log $g$} & \colhead{[Fe/H]} & \colhead{(km s$^{-1}$)} & \colhead{Observatory} \\
\colhead{(1)} & \colhead{(2)} & \colhead{(3)} & \colhead{(4)} & \colhead{(5)} & \colhead{(6)} & \colhead{(7)} & \colhead{(8)} & \colhead{(9)} & \colhead{(10)} & \colhead{(11)}}
\startdata

\input{tab1}

\enddata
\end{deluxetable}

\clearpage
\begin{deluxetable}{lc}
\tabletypesize{\scriptsize}
\tablewidth{0pc}
\tablecolumns{2}  
\tablecaption{Orbital Parameters for HD 167665 \label{orbital}}
\tablehead{
\colhead{Parameter} & \colhead{HD 167665}
}
\startdata

\input{tab2}

\enddata
\tablenotetext{a}{Uncertainty does not include contribution from host mass.}
\end{deluxetable}

\begin{deluxetable}{lcc}
\tabletypesize{\scriptsize}
\tablewidth{0pc}
\tablecolumns{3}
\tablecaption{Radial Velocities for HD 167665 \label{trverr167665}}
\tablehead{
\colhead{} & \colhead{RV} & \colhead{Uncertainty} \\
\colhead{JD-2,440,000} & \colhead{(\mps)} & \colhead{(\mps)}
}
\startdata
\input{tab3}
\enddata
\end{deluxetable}
\begin{deluxetable}{lcccccccc}
\tabletypesize{\scriptsize}
\tablewidth{0pc}
\tablecolumns{9}
\tablecaption{Constraints for Limited Phase Coverage Targets \label{limits}}
\tablehead{
\colhead{} & \multicolumn{2}{c}{$M$ sin $i$} & \colhead{} & \multicolumn{2}{c}{$P$} & \colhead{} & \multicolumn{2}{c}{$a$} \\
\colhead{} & \multicolumn{2}{c}{(\mjup)} & \colhead{} & \multicolumn{2}{c}{(yr)} & \colhead{} & \multicolumn{2}{c}{(AU)}  \\
\cline{2-3} \cline{5-6} \cline{8-9} \\
\colhead{Star} & \colhead{min} & \colhead{max} & \colhead{} & \colhead{min} & \colhead{max} & \colhead{} & \colhead{min} & \colhead{max}  \\
\colhead{(1)} & \colhead{(2)} & \colhead{(3)} & \colhead{} & \colhead{(4)} & \colhead{(5)} & \colhead{} & \colhead{(6)} & \colhead{(7)}
}
\startdata

\input{tab4}

\enddata
\end{deluxetable}

\end{document}

%% file: tab1.tex
142229  &  77810  &  G5V  &  40.5  &  5.04  &  1.06  &  5930  &  4.50  &  0.050  & 5.5 &  Keck  \\
150554  &  81662  &  F8V  &  44.8  &  4.43  &  1.07  &  6010  &  4.29  &  0.0  & 2.7 &  Keck  \\
167665  &  89620  &  G0V  &  29.7  &  4.04  &  1.09  &  6115  &  4.22  &  -0.17  & 5.2 & Keck  \\
211681  &  109169  &  G5 IV/V  &  70.9  &  3.85  &  1.31  &  5839  &  4.32  &  0.45  & 2.6 &  Keck  \\
215578  &  \nodata  &  K0 IV/V  &  43.2  &  4.33  &  1.04  &  5136  &  3.37  &  0.43  & 3.9 &  Keck  \\
217165  &  113438  &  G0V  &  43.7  &  4.47  &  1.05  &  5936  &  4.39  &  0.010  & 3.2 &  Keck  \\
29461  &  21654  &  G5 IV/V  &  48.1  &  4.53  &  1.10  &  5913  &  4.52  &  0.25  & 0.9 &  Keck  \\
31412  &  22919  &  F8V  &  36.0  &  4.24  &  1.13  &  6096  &  4.37  &  0.050  & 2.6 &  Keck  \\
5470  &  4423  &  G0V  &  67.8  &  4.19  &  1.13  &  5966  &  4.32  &  0.24  & 1.4 &  Keck  \\
8765  &  6712  &  G5 IV/V  &  75.1  &  3.77  &  1.04  &  5590  &  4.23  &  0.19  & 2.7 &  Keck  \\
199598  &  103455  &  G0V  &  33.2  &  4.32  &  1.11  &  6022  &  4.40  &  0.070  & 1.9 &  Lick  \\
72780  &  42112  &  F8V  &  52.5  &  3.87  &  1.23  &  6266  &  4.37  &  0.15  & 7.0 &  Lick  \\
73668  &  42488  &  G1V  &  35.7  &  4.50  &  1.05  &  5941  &  4.41  &  0.0  & 3.2 &  Lick  \\
74014  &  42634  &  K0V  &  33.5  &  4.96  &  0.968  &  5605  &  4.43  &  0.26  & 2.4 &  Lick  \\

%% file: tab2.tex
$P$ (days)  &  4385 $\pm$ 64  \\
$T_p$ (JD-2,440,000)  &  8140 $\pm$ 70  \\
$e$  &  0.337 $\pm$ 0.005  \\
$\omega$ (deg)  &  225.0 $\pm$ 1.1  \\
$K$ (\mps)  &  609.8 $\pm$ 3.1  \\
$a_{\rm min}$ (AU)  &  5.47 $\pm$ 0.05\tablenotemark{a}  \\
\msini\ (\mjup)  &  50.3 $\pm$ 0.4\tablenotemark{a}  \\
rms (\mps)  &  6.45  \\
$N_{obs}$  &  23  \\

%% file: tab3.tex
10283.918  &  436.87  &  2.10  \\
10604.041  &  329.14  &  1.67  \\
10665.871  &  295.83  &  1.84  \\
10955.098  &  170.89  &  2.29  \\
11013.928  &  151.26  &  2.50  \\
11069.828  &  101.16  &  2.10  \\
11313.053  &  -33.63  &  2.18  \\
11341.895  &  -41.21  &  1.95  \\
11367.816  &  -61.46  &  2.08  \\
11410.805  &  -100.66  &  2.57  \\
11439.715  &  -111.51  &  1.81  \\
11679.030  &  -281.80  &  1.86  \\
11702.962  &  -282.59  &  2.23  \\
12007.088  &  -522.40  &  2.00  \\
12095.949  &  -588.79  &  2.42  \\
12162.750  &  -599.04  &  2.09  \\
12390.040  &  -569.74  &  2.51  \\
12488.861  &  -482.93  &  2.36  \\
12574.701  &  -373.79  &  2.19  \\
13072.152  &  347.96  &  2.20  \\
13238.776  &  476.40  &  1.93  \\
13604.757  &  590.64  &  2.75  \\
13838.129  &  599.04  &  2.76  \\

%% file: tab4.tex
HD 142229 & 153 & \nodata &  & 16.4 & \nodata &  & 6.9 & \nodata  \\
HD 150554 & 65.7 & 75.2 &  & 8.0 & 33.7 &  & 4.2 & 11  \\
HD 211681 & 71.7 & 102 &  & 10.4 & 106 &  & 5.3 & 25  \\
HD 215578 & 523 & \nodata &  & 26.8 & \nodata &  & 10 & \nodata  \\
HD 217165 & 45.9 & \nodata &  & 11.1 & \nodata &  & 5.1 & \nodata  \\
HD 29461 & 91.2 & 224 &  & 10.2 & 25.9 &  & 5.0 & 9.6  \\
HD 31412 & 368 & 415 &  & 75.7 & 205 &  & 20 & 40  \\
HD 5470 & 163 & 425 &  & 13.0 & 167 &  & 6.0 & 35  \\
HD 8765 & 43.0 & 75.7 &  & 6.3 & 18.1 &  & 3.5 & 7.1  \\
HD 199598 & 105 & 120 &  & 25.6 & 85.3 &  & 9.3 & 21  \\
HD 72780 & 52.4 & 60.4 &  & 11.8 & 41.7 &  & 5.6 & 13  \\
HD 73668 & 160 & 181 &  & 17.1 & 24.1 &  & 7.1 & 8.9  \\
HD 74014 & 46.6 & 57.3 &  & 11.9 & 77.4 &  & 5.2 & 18  \\

%% file: ms.bbl
\clearpage
\begin{thebibliography}{}
\bibitem[Armitage \& Bonnell(2002)]{armitage02} Armitage, P. J., \& Bonnell, I. A. 2002, \mnras, 330, L11
\bibitem[Butler et al.(1996)]{butler96} Butler, R. P., Marcy, G. W., Williams, E., McCarthy, C., Dosanjh, P., \& Vogt, S. S. 1996, \pasp, 108, 500
\bibitem[Butler et al.(2006)]{butler06} Butler, R. P., et al.\ 2006, \apj, 646, 505
\bibitem[Burrows et al.(2006)]{burrows06} Burrows, A., Sudarsky, D., \& Hubeny, I.\ 2006, \apj, 640, 1063
\bibitem[Burrows et al.(1997)]{burrows97} Burrows, A., et al.\ 1997, \apj, 491, 856
\bibitem[Campbell et al.(1988)]{campbell88} Campbell, B., Walker, G.~A.~H., \& Yang, S.\ 1988, \apj, 331, 902
\bibitem[Cohen et al.(2003)]{cohen03} Cohen, M., Wheaton, W.~A., \& Megeath, S.~T.\ 2003, \aj, 126, 1090
\bibitem[Cox(2000)]{cox00} Cox, A. N. 2000, Allen's Astrophysical Quantities (4th ed.; New York: AIP)
\bibitem[Endl et al.(2004)]{endl04} Endl, M., Hatzes, A. P., Cochran, W. D., McArthur, B., Prieto, C. A., Paulson, D. B., Guenther, E., \& Bedalov, A. 2004, \apj, 611, 1121
\bibitem[ESA(1997)]{esa97} ESA 1997, VizieR Online Data Catalog, 1239, 0
\bibitem[Girardi et al.(2002)]{girardi02} Girardi, L., Bertelli, G., Bressan, A., Chiosi, C., Groenewegen, M.~A.~T., Marigo, P., Salasnich, B., \& Weiss, A.\ 2002, \aap, 391, 195
\bibitem[Grether \& Lineweaver(2006)]{grether06} Grether, D., Lineweaver, C. H. 2006, \apj, 640, 1051
\bibitem[Lindegren et al.(1997)]{lindegren97} Lindegren, L., et al.\ 1997, \aap, 323, L53
\bibitem[Marcy \& Butler(2000)]{marcy00} Marcy, G. W., Butler, R. P. 2000, \pasp, 112, 137
\bibitem[Marcy et al.(2005a)]{marcy05a} Marcy, G., Butler, R.~P., Fischer, D., Vogt, S., Wright, J.~T., Tinney, C.~G., \& Jones, H.~R.~A. 2005a, Progress of Theoretical Physics Supplement, 158, 24
\bibitem[Marcy et al.(2005b)]{marcy05b} Marcy, G. W., Butler, R. P., Vogt, S. S., Fischer, D. A., Henry, G. W., Laughlin, G., Wright, J. T., Johnson, \& J. A. 2005b, \apj, 619, 570
\bibitem[Matzner \& Levin(2005)]{matzner05} Matzner, C.~D., \& Levin, Y.\ 2005, \apj, 628, 817
\bibitem[McCarthy \& Zuckerman(2004)]{mccarthy04} McCarthy, C., \& Zuckerman, B.\ 2004, \aj, 127, 2871
\bibitem[Murdoch et al.(1993)]{murdoch93} Murdoch, K.~A., Hearnshaw, J.~B., \& Clark, M.\ 1993, \apj, 413, 349 
\bibitem[Nidever et al.(2002)]{nidever02} Nidever, D. L., Marcy, G. W., Butler, R. P., Fischer, D. A., \& Vogt, S. S. 2002, \apjs, 141, 503
\bibitem[Pourbaix \& Arenou(2001)]{pourbaix01} Pourbaix, D., \& Arenou, F. 2001, \aap, 372, 935
\bibitem[Pourbaix et al.(2004)]{pourbaix04} Pourbaix, D., et al. 2004, \aap, 424, 727
\bibitem[Reffert \& Quirrenbach(2006)]{reffert06} Reffert, S., \& Quirrenbach, A.\ 2006, \aap, 449, 699
\bibitem[Takeda et al.(2007)]{takeda07} Takeda, G., Ford, E.~B., 
Sills, A., Rasio, F.~A., Fischer, D.~A., \& Valenti, J.~A.\ 2007, \apjs, 
168, 297
\bibitem[Valenti \& Fischer(2005)]{valenti05} Valenti, J. A., \& Fischer, D. A. 2005, \apjs, 159, 141
\bibitem[Vogt(1987)]{vogt87} Vogt, S. S. 1987, \pasp, 99, 1214
\bibitem[Vogt et al.(1994)]{vogt94} Vogt, S. S., et al. 1994, \procspie, 2198, 362
\bibitem[Wright et al.(2004)]{wright04} Wright, J.~T., Marcy, G.~W., Butler, R.~P., \& Vogt, S.~S.\ 2004, \apjs, 152, 261
\bibitem[Wright(2005)]{wright05} Wright, J.~T.\ 2005, \pasp, 117, 657
\bibitem[Wright et al.(2007)]{wright07} Wright, J.~T., et al.\ 2007, \apj, in press.
\end{thebibliography}
